\documentclass{article}

\usepackage{PRIMEarxiv}

\usepackage[utf8]{inputenc} 
\usepackage[T1]{fontenc}    
\usepackage{url}            
\usepackage{booktabs}       
\usepackage{amsfonts}       
\usepackage{nicefrac}       
\usepackage{microtype}      
\usepackage{lipsum}
\usepackage{fancyhdr}       
\usepackage{graphicx}       
\graphicspath{{media/}}     

\usepackage{balance} 
\usepackage{etoolbox}
\usepackage[numbers]{natbib}
\usepackage{latexsym}
\usepackage{amssymb}
\usepackage{amsmath}
\usepackage{amsthm}
\usepackage{mathtools}
\usepackage{bm}
\usepackage{booktabs}
\usepackage{color}
\usepackage{multirow}
\usepackage{float}
\usepackage{adjustbox}
\usepackage{arydshln} 
\title{
		Computing Efficient Envy-Free Partial
		Allocations of Indivisible Goods}

\author{
  Robert Bredereck \\
  Institut f\"ur Informatik \\
  TU Clausthal \\
  Clausthal-Zellerfeld, Germany \\
  \texttt{robert.bredereck@tu-clausthal.de} \\
  \And
  Andrzej Kaczmarczyk \\
  Department of Computer Science \\
  The University of Chicago \\
  Chicago, USA \\
  \texttt{akaczmarczyk@uchicago.edu} \\
  \And
  Junjie Luo \\
  School of Mathematics and Statistics \\
  Beijing Jiaotong University \\
  Beijing, China \\
  \texttt{jjluo1@bjtu.edu.cn} \\
\And
  Bin Sun\\
  Institut f\"ur Informatik \\
  TU Clausthal \\
  Clausthal-Zellerfeld, Germany \\
  \texttt{bin.sun@tu-clausthal.de} \\
}

\newcommand{\BibTeX}{\rm B\kern-.05em{\sc i\kern-.025em b}\kern-.08em\TeX}

\usepackage{enumitem}

\newlist{custombul}{itemize}{1}
\setlist[custombul]{label=$\bullet$, labelindent=0pt, leftmargin=3ex}

\newtheorem{theorem}{Theorem}
\newtheorem{lemma}{Lemma}

\newcommand{\Agents}{\ensuremath{\mathcal{A}}}
\newcommand{\Observers}{\ensuremath{\mathcal{W}}}
\newcommand{\Resources}{\ensuremath{\mathcal{R}}}
\newcommand{\utility}{\ensuremath{u}}
\newcommand{\collect}[2]{\ensuremath{(#1_{#2})}}
\newcommand{\collectu}[1]{\collect{\utility}{#1}}
\newcommand{\collectua}{\collectu{a}}
\newcommand{\Forutils}{Regarding}
\newcommand{\forutils}{regarding}
\newcommand{\naturals}{\ensuremath{\mathbb{N}}}
\newcommand{\naturalsZero}{\ensuremath{\naturals_{0}}}
\newcommand{\allocation}{\ensuremath{\bm{\pi}}}

\newcommand{\allocbundleof}[1]{\ensuremath{\bm{\pi}(#1)}}
\newcommand{\Efficiency}{\ensuremath{\mathcal{E}}}
\newcommand{\threshold}{\ensuremath{t}}

\newcommand{\efpalongGen}[1]{{#1}-Envy-Free Partial Allocation}
\newcommand{\efpalongE}{\efpalongGen{\Efficiency}}
\newcommand{\efpashortGen}[1]{{#1}-EF-PA}
\newcommand{\efpashortE}{\efpashortGen{\Efficiency}}

\DeclareMathOperator{\usw}{usw}
\DeclareMathOperator{\esw}{esw}
\DeclareMathOperator{\mcar}{mcar}
\DeclareMathOperator{\size}{size}

\usepackage[textsize=scriptsize]{todonotes}

\usepackage{cleveref}
\crefname{obs}{observation}{observations}
\crefname{obs}{observation}{observations}
\Crefname{theorem}{thm.}{thms.}
\Crefname{theorem}{Thm.}{Thms.}
\Crefname{lemma}{lemma}{lemma.}
\Crefname{lemma}{Lemma}{Lemmas}

\newcommand{\appsymb}{$\bigstar$}
\newcommand{\appref}[1]{{\hyperref[proof:#1]{\appsymb}}}
\newcommand{\apprefX}[1]{{\hyperref[#1]{\appsymb}}}

\newcommand{\appendixsection}[1]{%
	\gappto{\appendixProofText}{\section{Additional Material for Section~\ref{#1}}\label{app:#1}}
}

\begin{document}

	
	\pagestyle{fancy}
	\fancyhead{}
	
	
	\maketitle

\begin{abstract}
	Envy-freeness is one of the most prominent fairness concepts in the allocation
	of indivisible goods. Even though trivial envy-free allocations always exist,
	rich literature shows this is not true when one additionally requires some
	efficiency concept (e.g., completeness, Pareto-efficiency, or social welfare maximization). In fact, in such case even deciding the existence of an efficient envy-free allocation is notoriously computationally hard. In this paper, we explore the limits of efficient computability by relaxing standard efficiency concepts and analyzing how this impacts the computational complexity of the respective problems. Specifically, we allow partial allocations (where not all goods are allocated) and impose only very mild efficiency constraints, such as ensuring each agent receives a bundle with positive utility. Surprisingly, even such seemingly weak efficiency requirements lead to a diverse computational complexity landscape. We identify several polynomial-time solvable or fixed-parameter tractable cases for binary utilities, yet we also find NP-hardness in very restricted scenarios involving ternary utilities.
\end{abstract}





	
	\maketitle 
	
	
	\section{Introduction}
	Computing fair allocations of indivisible resources is an important issue with
	many applications in all kinds of disciplines~\cite{BL08,BCM16,Wal15}.
	Envy-freeness, which ensures that no agent strictly prefers the resources allocated
	to a different agent over their own, is one of the most prominent fairness concepts~\cite{BCM16}.
	Unfortunately, non-trivial envy-free allocations do not always exist, and
	computing them is often associated to computationally very difficult problems~\cite{BL08}.
	In consequence, researchers have developed several ways to relax that fairness notion, such as envy-free up to one good (EF1)~\cite{Bud11} and envy-free up to any good (EFX)~\cite{CKMPSW19}.
	
	If one has a close look, however, then one quickly realizes that envy-freeness alone
	does not enforce any computational or existence issues: allocating no resource to anyone is envy-free.
	When adding an \emph{efficiency} component, such as requiring each resource to be
	allocated to someone (completeness), the picture changes.
	A folklore example is an instance with $n$~agents (say employees) and $n+1$~identical resources (say laptops):
	in every possible complete allocation there is at least one agent~$a$ who gets at most one resource and another
	agent~$a'$ that gets at least two resources, so that (for reasonable preferences) $a$~envies~$a'$.
	While there are certainly applications where this is indeed a problem, there is likely
	a trivial solution in most applications: allocating only $n$~of the $n+1$~resources (one to each agent).
	Such observations lead to the main question of our paper:
	which (weaker) efficiency concepts can help to identify additional (in comparison to completeness) envy-free allocations
	and what is the consequence on the computational complexity of finding such allocations?
	
	We come up with two basic ideas: What if the goal is not to allocate \emph{all resources}, but to
	either just allocate \emph{some resources} to the agents or just provide \emph{some utility} for the agents?
	In each case, we can focus on either the whole society or individual agents.
	More concretely, we ask for an envy-free (partial) allocation that
	(i) allocates at least $t$~resources in total, or
	(ii) allocates at least $t$~resources to each agent, or
	(iii) has utilitarian welfare of at least~$t$, or
	(iv) has egalitarian welfare of at least~$t$.
	
	Note that even variants for~$t=1$ have meaningful (potential) applications.
	They allow us to ask if there is an envy-free allocation of (some of) the resources such that
	(i) at least one resource is allocated,
	(ii) each agent gets at least one resource,
	(iii) at least one agent has a positive value for the allocated resources, or
	(iv) each agent has a positive value for the allocated resources.
	The first two cases (i,ii) model natural formal requirements while
	the other two cases (iii,iv) model basic (individual) quality requirements.
	
	The efficiency requirements are also relevant from the
	computational complexity perspective. To see this, assume---as we do in our
	paper---that the resources are goods, that is, agents report non-negative
	utilities for them. In this case, all our efficiency concepts for $t=1$ are
	significantly less demanding than multiple other prominent efficiency concepts, 
	such as completeness, as demonstrated by the earlier folklore example. 
	Hence, analyzing computational complexity 
	of these very special cases allows us to identify borders of
	efficient computability more accurately than before. On the other hand, if we
	find efficient algorithms for these relaxed cases, their results can be
	practically interpreted as the minimum efficiency levels that can be achieved.
	Indeed, given an instance of an allocation problem, by computing the result with
	such an algorithm, one can argue that any fair allocation that is less
	efficient is unjustified. Before we describe our findings, we briefly review the
	related literature to present the context helpful to interpret our results.
	
	\subsection{Related Work}
	
	
	Computing fair and efficient allocations has recently emerged as a very
	prominent stream of research in the area of fair allocation of indivisible
	resources. Allocations with maximum Nash welfare are both Pareto optimal and EF1, but computing such allocations is
	NP-hard~\cite{CKMPSW19}. Likewise, computing an allocation with the highest
	utilitarian social welfare among all EF1 allocations is NP-hard even for two
	agents~\cite{AzizHMS23}. As discussed before, the main difference of our model
	is that we allow partial allocations, and consequently we consider envy-freeness
	instead of its relaxation EF1.

	Allowing partial allocations is an important approach to guarantee the existence
	of EFX allocations (the existence of EFX complete allocations is still an open question).  
	\citet{DBLP:conf/ec/CaragiannisGH19} showed that there always exists an EFX
	partial allocation with at least half of the maximum Nash welfare.
	\citet{DBLP:journals/siamcomp/ChaudhuryKMS21} showed that donating at most $n-1$
	resources can guarantee the existence of EFX allocation such that no agent prefers the
	donated resources to its own bundle, where $n$ denotes the number of agents.
	This bound was latter improved to $n-2$ in general and to $1$ for the case with
	four agents~\cite{BergerCFF22}.
	Besides existence, \citet{Bu22} studied the problem of computing partial
	allocations with the maximum utilitarian welfare among all EFX allocations.
	Our work differs from this stream of research in that we focus on envy-freeness
	instead of EFX.
	
	\citet{AzizSW16} studied the problem of deleting (or adding) a minimum number of
	resources such that the resulting instance admits an envy-free allocation; which
	is equivalent to finding an envy-free allocation with the maximum size.
	However, they consider ordinal preferences whereas we consider cardinal preferences.
	Moreover, \citet{AzizSW16} considered the number of deleted resources, where the
	problem is NP-hard even if no resource can be deleted. In contrast, we consider
	the dual parameter the lower bound on the allocated resources to identify polynomial-time solvable cases.
	
	\citet{DBLP:journals/aamas/BoehmerBHKL24} studied the problem of transforming a given unfair allocation into an
	EF or EF1 allocation by donating few resources.
	In addition to upper bounds on the number of donated resources and the decrease
	on the utilitarian welfare, they also consider the lower bounds on the remaining
	allocated resources and the remaining utilitarian welfare.
	\citet[Chap 5]{DBLP:journals/algorithmica/DornHS21} studied the same problem but focused
	on a different fairness notion.
	The most prominent difference to our work is that in our model there is no given
	allocation.
	
	\citet{HosseiniSVWX20} introduced a fairness notion where agents can hide some
	of the resources in their own bundles such that no agent is envious assuming
	that the agents do not know the existence of the hidden resources in other agents'
	bundles.  Then the goal is to find a complete allocation and a minimum number of
	hidden resources such that no agent is envious. While the idea is similar to
	find an envy-free partial allocation with the maximum size, note that the hidden
	resources are not deleted; their owners get utility from them just like normal
	resources.
	
	A series of works \cite{GanSV19,BeynierCGHLMW19,KamiyamaMS21} studied the computational complexity of finding an envy-free house allocation when the number of houses is larger than the number of agents.
	This is equivalent to finding an envy-free (partial) allocation that allocates \emph{exactly} one resource to each agent. 
	Our model does not have this kind of upper bound on the number of resources allocated to each agent.
	\citet{aigner2022} studied the problem of finding an envy-free matching of maximum cardinality in a bipartite graph.
	Taking the bipartite graph as the representation of binary utilities of agents on one side towards resources on the other side, the problem studied by \citet{aigner2022} is equivalent to finding an envy-free (partial) allocation with the maximum size such that each agent gets at most one resource liked by it.
	Our model differs from it in that we do not add an upper bound for agents' bundles and we allow agents to receive resources with utility 0.
	Nevertheless, many of our algorithms for binary utilities use the structural
	properties of envy-free matchings 
	by~\citet{aigner2022}.

	
	\begin{table*}[t]
		\caption{Summary of results. Columns denote different utility constraints and
			efficiency threshold~$t$ values.
			Rows represent different efficiency concepts~$\Efficiency$. The hardness
			results for~$t=1$ apply to every positive~$t$ as well.} \label{tab:results}
		\centering
		\renewcommand{\arraystretch}{1.25}
		\setlength{\tabcolsep}{12pt}
		\begin{tabular}{rcccc}
			\toprule
			& Identical & \multicolumn{2}{c}{Binary} &	Ternary \\ 
			\cmidrule(lr){3-4}
			& $t=1$  & $t=1$ & $t$ & $t=1$ \\\midrule
			utilitarian social welfare ($\usw$) & \multirow{4}{*}{NP-h} & P & NP-h (FPT) & NP-h \\ 
			egalitarian social welfare ($\esw$) & ~ & P & P & NP-h \\ 
			\#resources allocated ($\size$) & ~ & P & NP-h (FPT) & NP-h  \\ 
			min-cardinality ($\mcar$) & ~ & P & NP-h & NP-h \\ \bottomrule
		\end{tabular}      
	\end{table*}

	\subsection{Contributions and Outline}
	
	We study the computational complexity of finding envy-free partial allocations
	with mild efficiency requirements.
	To this end, we consider a lower bound~$t$ on utilitarian welfare, egalitarian
	welfare, the number of allocated resources, or the minimum bundle size among all agents.
	Formal definitions can be found in \Cref{sec:pre}.
	An overview of our results is provided in \Cref{tab:results}.
	In \Cref{sec:identical}, we show that finding such allocations is strongly
	NP-hard, even if all agents have identical preferences.
	In \Cref{sec:binary}, we focus on the case with binary utilities, where each agent values a resource as either 0 or 1.
	We show that all the four variants are polynomial-time solvable when~$t=1$, indicating that determining the existence of envy-free allocations with minimal efficiency requirements can be done efficiently.
	For arbitrary $t$, while most problem variants become strongly NP-hard, we show that the utilitarian welfare variant and the number of allocated resources variant are both fixed-parameter tractable (FPT)\footnote{A problem is fixed-parameter tractable with respect to some parameter~$k$ if it can be solved in~$f(k)|I|^{O(1)}$~time, where $|I|$ denotes the input size.} with respect to $t$, implying that the problems can still be efficiently solved for small $t$.
	A surprising exception is the egalitarian welfare variant (which is typically harder than the utilitarian welfare):
	We show a polynomial-time algorithm that finds an envy-free partial allocation
	where each agent obtains
	a bundle with value at least~$t$ (for arbitrary~$t$).
	In \Cref{sec:ternary}, we go beyond binary preferences and allow for three different utility values.
	We show a reduction from the egalitarian welfare variant to the other three variants for ternary utilities and $t=1$, which reveals an interesting connection between the four efficiency requirements and might be of independent interest.
	Based on this reduction, we show that all variants become strongly NP-hard already when~$t=1$ for any ternary utility values $\{0,v,u\}$ with $0<v<u$.
Furthermore, all the problems shown to be NP-hard in this paper are contained in NP, since verifying that an allocation (guessed non-deterministically) is envy-free and meets the respective efficiency criterion is possible
		in polynomial time.
	
	\section{Preliminaries}\label{sec:pre}
	We fix a collection~$\Resources$ of $m$~\emph{resources}
	and a set~$\Agents$ of $n$~\emph{agents}.
	Each agent~$a \in \Agents$ reports its cardinal \emph{utility} from each resource via the
	\emph{utility function}~$\utility_a: \Resources \rightarrow \naturalsZero$.\footnote{\naturalsZero{} denotes the set of all non-negative integers.}
	We assume additive utilities, hence, with a slight abuse of notation,
	for some set~$B \subseteq \Resources$ of resources, the utility~$\utility_a(B)$
	of agent~$a \in \Agents$ from~$B$ is the sum of the agent's utilities for each
	resource in~$B$, i.e., $\utility_a(B) \coloneqq \sum_{r \in B} \utility_a(r)$.
	
	We use specific classes of cardinal utilities reported by agents.
	\emph{Identical utilities} denote a family of utilities in which every agent's
	utility functions are the same.
	The utilities are~\emph{binary} if agents' utilities
	use only values $0$ or~$1$ and~\emph{ternary} when
	there are three possible values of utility that agents can report.
	
	An \emph{allocation}~$\allocation{}: \Agents \rightarrow 2^{\Resources}$  
	assigns each agent~$a \in \Agents$ its private \emph{bundle}~$\allocbundleof{a}$, 
	i.e.,\ $\allocbundleof{a} \cap \allocbundleof{a'} = \emptyset$ for each distinct $a, a' \in \Agents$.
	If $\allocbundleof{i} = \emptyset$, it is an \emph{empty bundle}.
	If $\allocation$ is a partition of~$\Resources$, we say that
	$\allocation$ is \emph{complete}, otherwise we call it \emph{partial}.
	We call the smallest number~$\mcar(\allocation) \coloneqq \min_{a \in \Agents}
	|\allocbundleof{a}|$ of resources allocated to some agent the
	\emph{min-cardinality} of~$\allocation$, whereas by~$\size(\allocation)
	\coloneqq \sum_{a \in \Agents} |\allocbundleof{a}|$ we denote the total number of
	resources allocated by~$\allocation$.
	
	Given an allocation~$\allocation: \Agents \rightarrow 2^{\Resources}$ and some collection~$\collectua_{a \in \Agents}$
	of utility functions, we say that agent~$a \in \Agents$ is \emph{envious} \forutils~$\collectua$ under~$\allocation$
	if there is another agent~$a' \in \Agents$ whose
	bundle~$\allocbundleof{a'}$ is preferred by $a$ over their own
	bundle~$\allocbundleof{a}$; formally $\utility_a(\allocbundleof{a'}) >
	\utility_a(\allocbundleof{a})$.
	An allocation~$\allocation$ is \emph{envy-free} \forutils~$\collectua$ if no
	agent is envious under~$\allocation$.
	The \emph{utilitarian social welfare}~$\usw(\allocation)$ of~$\allocation$ \forutils~$\collectua$ is the sum of the utilities of
	agents for their bundles, i.e.,\ $\usw(\allocation) \coloneqq \sum_{a \in \Agents}
	\utility_a(\allocbundleof{a})$.
	Analogously, \emph{egalitarian social welfare}~$\esw(\allocation)$ is the minimum of the agent's utilities, i.e.,\
	$\esw(\allocation) \coloneqq \min_{a \in \Agents} \utility_a(\allocbundleof{a})$.
	(We omit ``\forutils~$\collectua$'' and ``under~$\allocation$'', respectively, when the context is clear.)
	
	Our problem of interest is a computational problem of deciding if, for a given
	input, one can find allocations that are envy-free and efficient. Following the
	introduction, we define our problem generally, using an efficiency measure
	placeholder~$\Efficiency$ to be substituted by any of the efficiency measures of
	our interest: utilitarian and egalitarian social welfare, size, and
	min-cardinality.
	
	\smallskip
	\hfill\begin{minipage}{\dimexpr\linewidth-1em}
		\noindent \textsc{\efpalongE{} (\efpashortE)}
		
		\noindent \textbf{Input:} A set~$\Resources$ of resources, 
		a set~$\Agents$ of agents, 
		a collection~$\collectua_{a \in \Agents}$
		of utility functions~$\utility_a\colon \Resources \rightarrow \naturalsZero$ 
		and an efficiency threshold~$\threshold$.
		
		\noindent \textbf{Question:} Is there an envy-free allocation~$\allocation$ such
		that~$\Efficiency(\allocation) \geq \threshold$?
	\end{minipage}


	\section{Identical valuations}
	\label{sec:identical}
	\appendixsection{sec:identical}  
	The case in which all agents have identical preferences is potentially simpler
	to solve than the general case when finding our desired allocations. However, we
	show even in this scenario, our problem is NP-hard for each efficiency notion we
	consider.
	
	We show hardness for all studied efficiency concepts with $t=1$ via a reduction from the \textsc{3-partition} problem~\cite{DBLP:books/fm/GareyJ79}.
	The main idea is to have one resource for each number of the \textsc{3-partition} instance as well as
	some well-designed dummy resources and extra agents, ensuring that each agent receives either one dummy
	resource or three non-dummy resources such that the utility for them adds up to the same value as
	the agents have for a dummy resource.

	\begin{theorem}\label{thm:identical-hard}
		For each~$\Efficiency \in \{\usw, \esw, \size, \mcar\}$ it holds that \efpashortE{} is strongly NP-hard,
		even if~$t=1$ and each agent has the same utility function.
	\end{theorem}

	{
		\begin{proof}
			The hardness proof proceeds by a reduction from the \textsc{3-partition}
			problem~\cite{DBLP:books/fm/GareyJ79}. Given a multiset of positive
			numbers~$N=\{e_1,e_2,\ldots,{e_{3n}}\}$, in \textsc{3-partition} we ask whether
			$N$~can be partitioned into $n$~three-element parts such that each of them sums
			up to $b \coloneqq \sum_{e_i \in N} e_i/n$.  Given an instance $\mathcal{I} = (N)$
			of~\textsc{3-partition}, we build an instance $ \mathcal{I'} = (\Resources,\Agents,\collectua,\threshold)$ of
			\efpashortE{}, with~$t=1$ and with identical utilities, as follows:
			\begin{custombul}
				\item We put $3n+1$~\emph{agents} $a_0$, $a_1$, $a_2$, $\ldots$, $a_{3n}$
				in \Agents;
				\item We construct $3n$~\emph{normal resources} $R_N=\{r_1,r_2,\ldots,r_{3n}\}$ and
				a set~$\Resources_S$ of \emph{special resources} with $|\Resources_S|=2n+1$, i.e.,  $\Resources=\Resources_N\cup \Resources_S$;
				\item We let each agent~$a \in \Agents$ have the same \emph{utility function}~$u$
				such that $u(r_i)=e_i+b$ for each resource $r_i\in \Resources_N$ and $u(r^*)=4b$ for each
				special resource $r^* \in \Resources_S$.
			\end{custombul}
			
			First we describe a structural property of every solution~$\allocation$ to
			$(\Resources,\Agents,\collectua)$. Note that it is clear that~$\allocation$ cannot be an
			empty allocation, as this would violate the threshold value~$t=1$ for each of
			efficiency concepts~$\Efficiency \in \{\usw, \esw, \size, \mcar\}$. Further,
			since there are $3n+1$~agents and only $3n$~normal resources, one agent has to get
			some special resource~$r^* \in \Resources_S$ consequently obtaining utility at least~$4b$.
			Otherwise, this agent would be envious. Hence, due to identical utilities, each
			agent has to get utility at least~$4b$ from their bundles. Finally, since the
			sum of the values of all resources is exactly $(3n+1)\cdot 4b$, it follows
			that~$\allocation$ assigns to each agent a bundle of value exactly~$4b$.
			Given that, it is easy to see that each agent that gets a special resource, does not
			have any other resource assigned by~$\allocation$. Thus, in~$\allocation$ there are
			exactly $2n+1$~agents that get one special resource. The remaining $n$~agents get
			exactly three normal resources out of~$3n$ of them. In any other case, because~$e_i
			< b$ for each $e_i \in N$, there would be at least one agent getting a utility
			smaller than~$4b$ from their bundle.
			
			We now show that the original instance~$\mathcal{I}$ is a YES-instance if and
			only if the constructed one, $\mathcal{I'}$, is a YES-instance. Our argument is
			solely based on the above-analyzed structure of solutions to~$\mathcal{I'}$.
			Since the structure applies to each studied efficiency concept when~$t=1$, the
			argument works for each choice of the efficiency concept.
			
			$(\Longrightarrow)$ Suppose $\mathcal{I}$ is a YES-instance of
			\textsc{3-partition}. Then, there are $n$ disjoint subsets, each summing up to $b$.
			We denote each subset $N_j=\{e_{x(j)},e_{y(j)}, e_{z(j)}\}$ for $j \in [n]$.
			We build allocation~$\allocation$ which certifies that $\mathcal{I'}$ is a
			YES-instance. First, $\allocation$ allocates the three resources $r_{x(j)},
			r_{y(j)}, r_{z(j)}$ to each agent $a_j$, $j\in [n]$. Then, it gives the
			remaining $2n+1$ special resources to the remaining $2n+1$~agents, one resource per
			agent. It is easy to verify that each agents gets utility exactly~$4b$ from
			their bundle, which shows that $\mathcal{I'}$ is indeed a YES-instance.
			
			$(\Longleftarrow)$ Let us assume that $\mathcal{I'}$ is a YES-instance. Due to
			the structure of solutions to~$\mathcal{I'}$ discussed earlier, we know that there is an
			allocation~$\allocation$ letting each agent have utility exactly $4b$.
			The structure also requires that~$\allocation$ assigns one special resource to
			$2n+1$~agents (one per agent) and that each of the remaining $n$~agents gets
			exactly three resources whose utility is~$4b$. Without loss of generality, we label
			the latter agents $a_1$, $a_2$, \ldots, $a_n$ and denote the resources of
			agent~$a_j$, $j \in [n]$, by $r_{x(j)}$, $r_{y(j)}$, and $r_{z(j)}$ for $j\in
			[n]$.  It follows that we can split~$N$ into~$n$ subsets constructing a
			subset~$N_j$ of numbers~$\{e_{x(j)},e_{y(j)}, e_{z(j)}\}$ for each $j\in [n]$.
			By our construction,
			the sum $e_{x(j)}+e_{y(j)}+e_{z(j)}$ of numbers in each such subset~$N_j$ is
			$b$, which proves that~$\mathcal{I}$ is a YES-instance.
			
			Clearly, the reduction is computable in polynomial time.
			
		\end{proof}
	}

	The presented result categorically
	sets the limits of our expectations, as the
	hardness holds for the weakest variants of efficiency concepts, that is, when
	the threshold~$t=1$. Hence, we focus on other aspects 
		to identify polynomial-time tractable cases.
	
	In the remaining sections, we will focus on restrictions on the set of utilities
	(resp.\ the images of the utility functions), since it seems essential that
	they are unrestricted in the above hardness reduction for identical preferences.
	Specifically, we study binary and ternary utilities, which are commonly studied 
	utility restrictions in the literature~\cite{BarmanKV18,DBLP:conf/wine/0002PP020,BabaioffEF21,DBLP:journals/corr/abs-2403-00943}.

	\section{Binary utilities}
	\label{sec:binary}
		
	Given that identifying exact utility values imposes a high cognitive
		burden for human agents, in practice binary utilities, where agents express preferences by pointing out
		which resources they desire and which not, are sometimes even
		preferred over more complicated variants. It is then easier to elicit correct
		preference data and to avoid excessive fatigue of the agents.
	
	The good news is that for binary preferences, our problem with $t=1$ is
	solvable in polynomial time for all the four efficiency notions. On the negative
	side, 
	for arbitrary~$t$, except for
	\efpashortGen{$\esw$}, the other three efficiency concepts yield NP-hardness.
	For some of these cases, however, we could find efficient (FPT) algorithms for bounded
	values of the threshold~$t$.
	
	
	\subsection{Egalitarian social welfare}
	Beginning with \efpashortGen{$\esw$}, we show that it is polynomial-time solvable
	by providing a reduction to computing a maximum cardinality matching in
	bipartite graphs.
	\begin{theorem}\label{thm:binary-esw}
		For $0/1$-utilities \efpashortGen{$\esw$} is solvable in $O(m^{2.5})$ time.
	\end{theorem}
	\begin{proof}
		If $t>\frac{m}{n}$, then no allocation can get $\esw({\allocation})\geq t$.
		So, in the following we assume $t \leq \frac{m}{n}$.
		Given an envy-free allocation~$\allocation$ with~$\esw(\allocation)\geq t$, we construct a new allocation~$\allocation'$ by keeping $t$ arbitrary resources from each agent's bundle that are liked by the agent and deleting the other resources. 
		Note that~$\allocation'$ also satisfies envy-freeness and~$\esw(\allocation')\geq t$.
		Therefore, it suffices to check whether there exists an allocation such that every agent gets exactly~$t$ resources liked by it.
		To this end, we create a bipartite graph where one side consists of~$t$ copies of each agent and the other side consists of all resources, and there is an edge between an agent and a resource if the agent likes the resource.
		Then there exists an envy-free allocation with~$\esw(\allocation)\geq t$ if and only if a maximum cardinality matching of this bipartite graph, which can be computed in~$O((tn)^{1.5}m)=O(m^{2.5})$~time~\cite{ramshaw2012minimum}, saturates the agent side.
	\end{proof}
	
	\subsection{Utilizing envy-free matchings}
	
	For the other three efficiency measures, we create a bipartite graph $G=(X
	\dot{\cup} Y,E)$, where $X=\Agents$, $Y=\Resources$, and there is an edge
	between $x_i \in X$ and $y_j \in Y$ if $u_i(r_j)=1$.
	We use the concept of \emph{envy-free matchings} (EFM) for bipartite
	graphs introduced by~\citet{aigner2022}.
	A matching $M$ in a bipartite graph $G=(X \dot{\cup} Y,E)$ is
	envy-free with respect to $X$ if no vertex in $X \setminus X_M$ is adjacent to
	any vertex in $Y_M$, where $X_M$ (resp. $Y_M$) represents the set of vertices
	from $X$ (resp. $Y$) saturated by $M$. Note that each envy-free matching~$M$ in
	$G=(X \dot{\cup} Y,E)$ induces an envy-free allocation~$\allocation^M$, where
	every agent gets at most one resource.
	Slightly abusing the notation, we sometimes use subsets of~$X$ (resp.\ $Y$) to
	denote the corresponding subsets of agents (resp.\ resources). 
	
	\citet{aigner2022} show that finding an envy-free matching of maximum cardinality is solvable in polynomial time. 
	The idea is to first compute an arbitrary matching~$M$ of maximum cardinality.
	Then, starting with each vertex from $X$ that is not saturated by $M$, we find $M$-alternating paths, which partition the vertex set into two parts according to whether they are covered by these paths or not.
	It is shown that this partition is independent of the initial matching $M$ and that all envy-free matchings are contained in the part not covered by the above $M$-alternating paths.
	In the following theorem, we
	summarize the findings of~\citet{aigner2022} related to envy-free matchings that
	are relevant to our results.
	
	\begin{theorem}[\cite{aigner2022}]
		\label{thm:EFM}
		Every bipartite graph $G=(X \dot{\cup} Y,E)$ admits a unique partition $X=X_S \dot{\cup} X_L$
		and $Y=Y_S \dot{\cup} Y_L$, called the \emph{EFM partition} of $G$, satisfying the following conditions:
		\begin{custombul}
			\item {An $X_L$-saturating matching in $G[X_L; Y_L]$ always exists,} and every $X_L$-saturating matching in $G[X_L; Y_L]$ is an envy-free matching in $G$;
			\item Every envy-free matching in $G$ is contained in $G[X_L; Y_L]$;
			\item There are no edges between $X_S$ and $Y_L$;
			\item Each vertex in $Y_S$ is connected to at least one vertex in $X_S$.
		\end{custombul}
		Moreover, the unique EFM partition and a maximum envy-free matching ($X_L$-saturating matching in $G[X_L; Y_L]$) can be computed in $O(|E|\sqrt{\min\{|X|,|Y|\}})$ time.
	\end{theorem}
	
	Based on \Cref{thm:EFM}, we derive the following lemma, which will be useful for designing algorithms in the remainder of this section.
	
	\begin{lemma}\label{lem:EF-property}
		For any envy-free allocation, all agents in~$X_S$ receive a bundle of
		utility~$0$ and all the allocated resources are from $Y_L$. 
	\end{lemma}
	
	\begin{proof}
		Given any envy-free allocation~$\allocation$, denote by~$\Agents_z$ the set of agents receiving a bundle of utility~$0$ and by~$\Agents_p$ the set of remaining agents
		(receiving a bundle of utility larger than~$0$).
		We construct a new allocation~$\allocation'$ as follows.
		For each agent from~$\Agents_z$, delete all resources from its bundle.
		For each agent from~$\Agents_p$ keep an arbitrary resource in its bundle with utility 1 for the agent and delete the other resources.
		We show that~$\allocation'$ is still envy-free.
		Since the original allocation~$\allocation$ is envy-free and all agents from~$\Agents_z$
		receive a bundle of utility~$0$ under~$\allocation$, it must be that every agent from~$\Agents_z$ values every resource allocated under~$\allocation$ as~$0$,
		and hence no agent from~$\Agents_z$ will envy other agents under~$\allocation'$.
		Moreover, under~$\allocation'$, every agent from~$\Agents_p$ receives a bundle of utility~$1$ and every agent gets exactly one resource,
		so no agent from~$\Agents_p$ will be envious.
		Therefore, $\allocation'$ is envy-free.
		Since each agent either gets nothing or gets one resource liked by it under~$\allocation'$, it induces an envy-free matching~$M$ in~$G=(X \dot{\cup} Y,E)$.
		According to \Cref{thm:EFM}, we have~$\Agents_p \subseteq X_L$.
		Since~$\Agents = X_S \cup X_L = \Agents_z \cup \Agents_p$, we have~$X_S \subseteq \Agents_z$, which means that all agents from~$X_S$ receive a bundle of utility~$0$.
		Since~$\allocation$ is envy-free, it follows that all the allocated resources under~$\allocation$ have utility~$0$ for agents from~$X_S$.
		According to \Cref{thm:EFM}, each resource in~$Y_S$ is liked by at least one agent from~$X_S$, so all the allocated resources are from~$Y_L$. 
	\end{proof}

	\subsection{Social welfare and allocation size}
	
	Based on \Cref{lem:EF-property}, we can design an FPT algorithm for \efpashortGen{$\usw$}.
	The idea is that according to \Cref{lem:EF-property}, it suffices to consider allocations restricted to $X_L$ and $Y_L$. If $|X_L| \geq t$, there is a trivial solution following from the envy-free matching. Otherwise, we can bound the size of the instance by a function depending only on $t$.
	
	\begin{theorem}\label{thm:binary-usw}
		For $0/1$-utilities \efpashortGen{$\usw$} is NP-hard and fixed-parameter tractable with respect to~$t$.
		In particular, if~$t=1$, then \efpashortGen{$\usw$} is solvable in~$O(n^{1.5}m)$~time for $0/1$-utilities.
	\end{theorem}
	
	\begin{proof}
		Hardness follows from the equivalence of \efpashortGen{$\usw$} for
		$0/1$-utilities with $t$ setting as the maximum utilitarian social welfare
		among all allocations and the NP-hard problem of deciding the existence of a
		Pareto efficient and envy-free allocation~\cite{BL08}, since \citet[Ob.1]{DBLP:conf/ijcai/BliemBN16} shows that, in case of $0/1$-utilities, an allocation is Pareto-efficient if and only if it is complete and every resource is allocated to an agent that assigns 1 to it.
		
		Next, we show that \efpashortGen{$\usw$} for $0/1$-utilities is fixed-parameter tractable with respect to~$t$.
		According to \Cref{lem:EF-property}, it suffices to check allocations that only allocate resources from~$Y_L$.
		In addition, since in any desired allocation agents from $X_S$ receive a bundle of utility~$0$, it suffices to check allocations that only allocate resources from~$Y_L$ to agents from~$X_L$.
		If~$X_L=\emptyset$, then no such allocations exists.
		In the following analysis we assume~$X_L \neq \emptyset$.
		According to \Cref{thm:EFM}, there exists an envy-free matching~$M$ of cardinality~$|X_L|$ in~$G[X_L; Y_L]$.
		If~$|X_L|\geq t$, then~$M$ induces an envy-free allocation with social welfare at least~$t$ and we are done.
		Otherwise, we have~$|X_L| < t$.
		Since agents have binary utilities, we can partition all resources from~$Y_L$ into at most~$2^{|X_L|} < 2^t$ groups according to the subset of agents from~$X_L$ who like the resource.
		If there is a group with more than~$t^2$ resources, then allocating each agent from~$X_L$ a different set of~$t$ resources from this group is an envy-free allocation with social welfare $t|X_L|\geq t$ and we are done. {This is because resources with zero utility for all agents are irrelevant and can be removed during preprocessing. Thus, every resource has a positive value for at least one agent.}
		Otherwise, we have~$|Y_L| < 2^tt^2$ and then we can bound the number of all possible allocations restricted to~$X_L$ and~$Y_L$ by~$O(2^{t^2}t^{2t})$.
		Thus, the problem is fixed-parameter {tractable for~$t$.}
		
		When~$t=1$, it suffices to compute the EFM partition
		of~$G$ and check whether~$|X_L|\geq 1$,
		so the running time is~$O(n^{1.5}m)$ according to \Cref{thm:EFM}.
	\end{proof}
	
	Next, we provide an FPT algorithm for \efpashortGen{$\size$} using similar ideas.
	Here we just need to consider allocations restricted to $Y_L$ and we will compare the size of $X$ (instead of $X_L$) and $t$.
	
	\begin{theorem}\label{thm:binary-size}
		For $0/1$-utilities \efpashortGen{$\size$} is NP-hard and is fixed-parameter tractable with respect to~$t$.
		In particular, if~$t=1$, then \efpashortGen{$\size$} is solvable in~$O(n^{1.5}m)$~time for $0/1$-utilities.
	\end{theorem}
	
	\begin{proof}
		{Hardness follows from \efpashortGen{$\size$} for $0/1$-utilities with $t=|\Resources|$ being
			equivalent} to the problem of deciding the existence of a complete and envy-free allocation, which is NP-hard~\cite{HosseiniSVWX20,AzizGMW15}.
		
		Next we show that \efpashortGen{$\size$} for $0/1$-utilities is fixed-parameter tractable with respect to~$t$.
		By \Cref{lem:EF-property}, it suffices to check allocations that only allocate resources from~$Y_L$.
		If~$|Y_L|<t$, then there is no such allocation with size at least~$t$.
		In the following analysis we assume~$|Y_L|>t$.
		If~$|X| \leq t$, then similar to the case for~$\usw$, we can bound the number of all possible allocations restricted to~$Y_L$ by~$O(2^{t^2}t^{2t})$, and hence the problem is fixed-parameter tractable with respect to~$t$.
		If~$|X| >t$, then we can find an envy-free allocation with size at least~$t$ as follows.
		According to \Cref{thm:EFM}, there exists an envy-free matching~$M$ of cardinality~$|X_L|$ in~$G[X_L; Y_L]$, which induces an envy-free allocation~$\allocation^M$.
		We extend~$\allocation^M$ by letting each agent from~$X_S$ select a different resource from~$Y_L \setminus Y_M$ until there is no remaining resource or each agent from~$X_S$ gets one resource. 
		Denote the resulting allocation by~$\allocation$.
		We have~$\size(\allocation)\geq \min\{|X|,|Y_L|\} \geq t$.
		According to \Cref{thm:EFM}, no resource from~$Y_L$ is liked by any agent from~$X_S$, so~$\allocation$ is still envy-free.
		
		For~$t=1$, computing the EFM partition of~$G$ and checking
		whether~$|Y_L|\geq 1$ suffices; so~\Cref{thm:EFM} yields running time~$O(n^{1.5}m)$.
	\end{proof}
	
	\subsection{Min-cardinality}
	Finally, we consider \efpashortGen{$\mcar$}.
	The following lemma reduces \efpashortGen{$\mcar$} with $t=1$ to comparing the cardinality of $X$ and $Y_L$ in the EFM partition of $G$.
	
	\begin{lemma}\label{lem:NEB=1}
		The following three statements are equivalent:
		\begin{enumerate}
			\item There exists an envy-free allocation~$\allocation$ where every agent gets a non-empty bundle, i.e., $\mcar(\allocation)\geq1$;
			\item There exists an envy-free allocation~$\allocation$ where every agent gets exactly one resource, i.e., $|\allocbundleof{a}|=1$ for each~$a \in \Agents$;
			\item $|X| \leq |Y_L|$.
		\end{enumerate}
	\end{lemma}
	
	\begin{proof}
		$(1) \Leftarrow (2)$:
		If there exists an envy-free allocation~$\allocation$ with~$|\allocbundleof{a}|=1$ for each~$a \in \Agents$, then clearly $\mcar(\allocation)\geq1$.
		
		$(1) \Rightarrow (2)$:
		Given an envy-free allocation~$\allocation$ with $\mcar(\allocation)\geq1$, denote by $\Agents_z$ the set of agents receiving a bundle of utility~$0$
		and by $\Agents_p$ the set of remaining agents (receiving a bundle of utility larger than~$0$).
		For each agent from~$\Agents_z$, keep an arbitrary resource in its bundle and delete the other resources.
		For each agent from~$\Agents_p$, keep an arbitrary resource in its bundle with utility 1 for the agent and delete the other resources.
		Denote by $\allocation'$ the resulting allocation, where every agent gets exactly one resource.
		It remains to show that $\allocation'$ is envy-free.
		Since the original allocation $\allocation$ satisfies envy-freeness and all agents from $\Agents_z$ have utility 0 under $\allocation$, it must be that every agent from $\Agents_z$ values every resource allocated under $\allocation$ as 0, and hence no agent from $\Agents_z$ will envy other agents under~$\allocation'$.
		Moreover, under $\allocation'$, since every agent from $\Agents_p$ has utility 1 and every agent gets exactly one resource, no agent from $\Agents_p$ will be envious.
		Thus, $\allocation'$ satisfies envy-freeness.  
		
		$(2) \Leftarrow (3)$:
		Suppose that $|X| \leq |Y_L|$. 
		According to \Cref{thm:EFM} we can find a $X_L$-saturating envy-free matching $M$ in $G[X_L; Y_L]$, which induces an envy-free allocation~$\allocation^M$, where every agent gets at most one resource.
		To get an envy-free allocation where every agent gets exactly one resource, we let each remaining agent corresponding to $X_S$ select a different resource from $Y_L \setminus Y_M$.
		Since $|Y_L| \geq |X|$, there are enough remaining resources from $Y_L \setminus Y_M$.
		Denote the resulting allocation by $\allocation$, where every agent now gets exactly one resource.
		Since there are no edges between $X_S$ and $Y_L$, all agents corresponding to $X_S$ are non-envious.
		For agents corresponding to $X_L$, since they all have utility 1 and every agent gets exactly one resource, all of them are non-envious. 
		Therefore, $\allocation$ is envy-free.
		
		$(2) \Rightarrow (3)$:
		Let~$\allocation$ be an envy-free allocation where every agent gets exactly one resource.
		According to \Cref{lem:EF-property}, all the allocated resources are from $Y_L$.
		Thus, $|X| \leq |Y_L|$. 
	\end{proof}
	
	It immediately follows that \efpashortGen{$\mcar$} with $t=1$ is solvable in
	polynomial time.  We subsequently prove the NP-hardness for the general case
	with arbitrary $t$. However, whether the problem is fixed-parameter tractable
	with respect to $t$ remains open.
	
	\begin{theorem}\label{thm:binary-mcar}
		For $0/1$-utilities \efpashortGen{$\mcar$} is NP-hard. 
		If~$t=1$ then it is solvable in~$O(n^{1.5}m)$~time.
	\end{theorem}
	
	\begin{proof}
		We show the NP-hardness of \efpashortGen{$\mcar$} by providing a simple
		many-one reduction from \efpashortGen{$\size$} with $t=|\Resources|$, which
		is shown to be NP-hard in \Cref{thm:binary-size}. 
		Given an instance $(\Agents, \Resources, t=|\Resources|)$ of \efpashortGen{$\size$}, we create an instance $(\Agents, \Resources', t')$ of \efpashortGen{$\mcar$}, where $\Resources'$ contains all resources in $\Resources$ and also $t(|\Agents|-1)$ dummy resources that are not liked by any agent, and $t'=t$. 
		It is easy to verify that there exists an envy-free and complete allocation for the former instance if and only if there exists an envy-free allocation such that every agent gets exactly $t$ resources for the latter instance.
		
		When $t=1$, according to \Cref{lem:NEB=1} and \Cref{thm:EFM}, it suffices to compute the EFM partition for $G$ and check whether $|X| \leq |Y_L|$, so the running time is~$O(n^{1.5}m)$.
	\end{proof}

	\section{Ternary Valuations}
	\label{sec:ternary}
	\appendixsection{sec:ternary}
	
	We have seen that our problems are tractable for binary preferences and~$t=1$,
	which already has quite clear practical relevance as discussed in the introduction.
	A very natural question is whether these positive results transfer to three different utility values. 
	In this section we answer this question negatively by showing strong NP-hardness
	for all the four goals under any three different utility values $\{0,v,u\}$ with $0<v<u$. 
	
		
		
	We start by providing a very general reduction from $\esw$ to the other three problems for any ternary utilities which include
	utility zero and $t=1$.
	
	\begin{lemma} 
		\label{lemma:ter+usw+mcar+size}
		Let~$v$ and~$u$ be two positive integers with $0<v<u$.
		Let $\Resources$~be a set of resources,
		$\Agents$~be a set of agents,
		and $\collectua_{a \in \Agents}$ be a collection of utility functions with $u_a: \Resources \rightarrow \{0,v,u\}$.
		Then, there exist extended sets of resources~$\Resources^*=\Resources \cup \Resources_{\text{shadow}}$
		and agents~$\Agents^*=\Agents \cup \Agents_{\text{shadow}}$,
		and a collection of extended utility functions $\collect{u^*}{a}_{a \in A^*}$ (with $u^*_{a}(r) = u_{a}(r)$ for each $a\in \Agents$ and each $r \in \Resources$)
		such that:
		
			\Forutils~\collectua{} there exists an envy-free allocation $\pi^{\esw}:\Agents \rightarrow 2^{\Resources}$ with $\esw(\pi^{\esw})\geq 1$, if and only if
			\forutils~\collect{u^*}{a} there exists an envy-free allocation $\pi^*:\Agents^* \rightarrow 2^{\Resources^*}$ with  $\Efficiency(\pi^*)\geq 1$ for each $\Efficiency \in \{\mcar,\usw,\size\}$\footnote{Note that given any $\pi^{\Efficiency}$ for $\Efficiency \in \{\esw, \mcar,\usw,\size\}$,
				we can compute each of the respective other allocations in polynomial time. Here, the condition $\Efficiency(\pi^*)\geq 1$ corresponds to the setting $t=1$ in \efpashortE.}.
		Moreover, $(\Resources^*,\Agents^*,(u^*_a)_{a \in \Agents^*})$ can be computed in linear time. 
	\end{lemma}

	\begin{proof}
		Given $(\Resources,\Agents,(u_a)_{a\in \Agents})$, we construct $(\Resources^*=\Resources \cup \Resources_{\text{shadow}}, \allowbreak \Agents^*=\Agents \cup \Agents_{\text{shadow}},(u^*_a)_{a\in A^*})$ as follows.
		For each resource, we create two corresponding shadow agents and two corresponding shadow resources. That is, $\Agents_{\text{shadow}}:=\{a'_r, a''_r \mid r \in R\}$ and $\Resources_{\text{shadow}}  \allowbreak  :=\{r', r'' \mid r \in \Resources\}$.
		We distinguish between \emph{original agents}~$\Agents$ and \emph{shadow agents}~$\Agents_{\text{shadow}}$, as well as between \emph{original resources}~$\Resources$ and \emph{shadow resources}~$\Resources_{\text{shadow}}$.
		The idea is to define utilities functions $(u^*_a)_{a\in A^*}$ such that  whenever any agent gets a resource, each shadow agent will also require a shadow resource, which in turn ensures that every agent gets a resource of positive value.
		Formally, $(u^*_a)_{a\in A^*}$ is defined as follows (see also Table~\ref{tab:ter+usw+mcar+size}).
		\begin{custombul}
			\item For each original agent $a$ and each original resource $r$, $u^*$ is
			identical to~$u$, i.e., $u^*_{a}(r)=u_{a}(r)$. 
			\item Each original agent is interested in all the shadow resources and values each of them as~$v$. 
			\item Each shadow agent is interested in all the shadow resources and values each of them as~$u$.
			\item Each shadow agent ${a'_r}$ or ${a''_r}\in \Agents^*_{\text{shadow}}$ is also interested in its unique corresponding original resource $r\in \Resources$, i.e., $u^*_{a'_r}(r)=u^*_{a''_r}(r)=v$, and values all other original resources as~$0$.
		\end{custombul} 
		{
			\renewcommand{\arraystretch}{1.25}
			\setlength{\tabcolsep}{8.5pt}
			\begin{table}[t]
				\caption{Agent's utility functions in the proof of~\Cref{lemma:ter+usw+mcar+size}.}
				\label{tab:ter+usw+mcar+size}
				\centering
				\begin{tabular}{c|ccc}
					& $\bar{r}\in \Resources\setminus\{r\}$ & $r\in \Resources$ & $r^*\in \Resources_{\text {shadow}}$ \\
					\hline
					$a\in \Agents$ & $u_{a}(\bar{r})$ & $u_{a}(r)$ & $v$ \\
					${a'_r},{a''_r}\in \Agents_{\text{shadow}}$ & 0 & $v$ & $u$ \\
				\end{tabular}
			\end{table}
		}
		
		Next, we show that for $(\Resources^*,\Agents^*,(u^*_a)_{a \in \Agents^*})$ and any $\Efficiency,\Efficiency' \in \{\mcar,\usw,\size\}$
		it holds that for every envy-free allocation $\pi$ with $\Efficiency(\pi)\geq 1$ we also have $\Efficiency'(\pi)\geq 1$.
		By definition, it is obvious that an envy-free allocation~$\pi$ with $\mcar(\pi)\ge1$ or $\usw(\pi)\ge1$
		must in both cases have $\size(\pi)\ge1$.
		Let us conversely assume that there exists some envy-free allocation $\pi$ with $\size(\pi)\ge1$ for $(\Resources^*,\Agents^*,(u^*_a)_{a \in \Agents^*})$.
		We want to show that $\mcar(\pi)\ge1$ and $\usw(\pi)\ge1$ also hold for $(\Resources^*,\Agents^*,(u^*_a)_{a \in \Agents^*})$.
		Since $\size(\pi) \ge 1$, at least one resource~$r$ is allocated. If $r$ is not a shadow resource, then at least one of the two corresponding shadow agents~$a'_r$ or~$a''_r$
		gets a shadow resource. Thus, at least one shadow resource is allocated under~$\pi$. 
		Considering that each shadow agent can only gain a maximum value of~$v$ from the original resources, and $u>v$, the fact that at least one shadow resource is allocated under~$\pi$ makes
		every shadow agent require at least one shadow resource with value at least $u$.
		Since $|\Agents_{\text{shadow}}|=|\Resources_{\text{shadow}}|=2|\Resources|$, each shadow agent should receive exactly one shadow resource.
		Since each original agent values each shadow resource as $v$, this enforces that each original agent gets a bundle with value at least $v$.
		Therefore, we have $\mcar(\pi)\ge1$ and $\usw(\pi)\ge1$.

		To prove the lemma, it remains to show that there exists an envy-free allocation $\pi^{\esw}$ with $\esw(\pi)\geq 1$ for $(\Resources,\Agents,(u_a)_{a \in \Agents})$ if and only if 
		there exists an envy-free allocation $\pi^{\size}$ with $\Efficiency(\pi^{\size})\geq 1$ for $(\Resources^*, \Agents^*,(u^*_a)_{a \in \Agents^*})$.
		
		$(\Longrightarrow)$ Assume there exists an envy-free allocation $\pi^{\esw}$ with $\esw(\pi^{\esw})\geq 1$ for $(\Resources,\Agents,(u_a)_{a \in \Agents})$.
		We construct a desired allocation~$\pi^{\size}$ for
		$(\Resources^*,\Agents^*,(u^*_a)_{a \in \Agents^*})$ as follows.
		Analogously to~$\pi^{\esw}$, we let~$\pi^{\size}_a=\pi^{\esw}_a$ for each original agent~$a \in \Agents$.
		Aside from that, each shadow agent is assigned an arbitrary shadow resource.
		Clearly, original agents will not envy each other, and each of them receives a bundle with positive value of at least $v$.
		Consequently, original agents will not envy shadow agents either, since they perceive the value of each shadow agent's bundle to be exactly $v$.
		Meanwhile, shadow agents will not envy original agents because, in their views, the value of each shadow agent's bundle is~$u$, whereas the value of any original agent's bundle does not exceed~$v$. 
		
		$(\Longleftarrow)$ Assume there exists some envy-free allocation~$\pi^{\size}$ with $\size(\pi^{\size})\ge1$ for $(\Resources^*,\Agents^*,(u^*_a)_{a \in \Agents^*})$.
		Recall that in $\pi^{\size}$, each shadow agent must get exactly one shadow
		resource, and each original agent must get a bundle with a positive value.
		Thus, we have $\esw(\pi^{\size})\ge1$.
		We create an allocation~$\pi^{\esw}$ for $(\Resources,\Agents,(u_a)_{a \in \Agents})$ in a straight-forward way by setting
		$\pi^{\esw}_a:=\pi^{\size}_a$ for each original agent~$a \in A$.
		Note that this is indeed a well-defined allocation for~$(\Resources,\Agents,(u_a)_{a \in \Agents})$ since~$\pi^{\size}$ allocates shadow resources only to shadow agents.
		Since the original agents do not envy each another in~$\pi^{\size}$ for $(\Resources^*,\Agents^*,(u^*_a)_{a \in \Agents^*})$, and the utility functions of the original agents for original resources are identical for $(\Resources^*,\Agents^*, \allowbreak (u^*_a)_{a \in \Agents^*})$ and $(\Resources,\Agents,(u_a)_{a \in \Agents})$, it follows that $\pi^{\esw}$~is envy-free for $(\Resources,\Agents,(u_a)_{a \in \Agents})$.
	\end{proof}
	
	According to \Cref{lemma:ter+usw+mcar+size}, if we show that \efpashortGen{$\esw$} is strongly NP-hard for ternary utility values $0<v<u$,
	then we automatically also get the strong NP-hardness of
	\efpashortGen{$\Efficiency$} for each $\Efficiency  \in \{\mcar,\usw, \size\}$.
	Our main result in this section is that all the four goals are strongly NP-hard for ternary utility values $0<v<u$ even if $t=1$, stated as follows.
	


	\begin{theorem} \label{thm:ter}
	Let $\Efficiency  \in \{\esw,\mcar,\usw, \size\}$ and let
			$v,u \in \mathbb{N}$ be fixed with $0<v<u$.
			Then, \efpashortE{} is strongly NP-hard,
		even if each agent assigns only values from~$\{0,v,u\}$ to the resources and~$t=1$.
	\end{theorem}

	By \Cref{lemma:ter+usw+mcar+size}, it suffices to show the strong
	NP-hardness for \efpashortGen{$\esw$}.
	To this end, the following \Cref{lemma:ter+usw+mcar+size} to
	\ref{lem:ter-v-kv+c} serve as a case distinction over the values of~$u$ and~$v$.
	Each lemma shows a different reduction from the NP-hard \textsc{Exact Cover by
		3-Sets (X3C)} problem~\cite{DBLP:books/fm/GareyJ79}. Given a multiset
	$X=\{x_1,x_2,\ldots,x_{3n}\}$ and a collection $C=\{S_1,S_2,\ldots,S_{m}\}$ of
	3-element subsets of $X$, \textsc{X3C} asks whether there is some $C' \subseteq
	C$ where every element of $X$ occurs in exactly one member of $C'$.  We assume
	without loss of generality that~$m>3n$, as we can always add dummy $3$-sets to
	guarantee this.
	
	
	\begin{lemma} \label{lem:ter-1-3}
		\efpashortGen{$\esw$} with ternary utility values $\{0,v,u\}$, $u=kv>0$, $k\geq 3$, and $t=1$  is strongly NP-hard.
	\end{lemma}
	
	\begin{proof}
				The hardness proof proceeds by a reduction from X3C.
				Given an instance $(X,C)$ of the \textsc{X3C}, we construct an instance $\mathcal{I} = (\Resources,\Agents,(u_a)_{a\in \Agents},t=1)$ of the \efpashortGen{$\esw$} problem as follows.
				\begin{custombul}
					\item There are $m$ \emph{cover agents} $\Agents_C=\{a_1, a_2, \ldots, a_m\}$ and a \emph{special agent} $a^*$, i.e., $\Agents=\Agents_C\cup \{a^*\}$.
					\item There are $3n$ \emph{normal resources} $\Resources_N=\{r_1,r_2,\ldots,r_{3n}\}$, $(k-3)n$ \emph{small resources}
					$\Resources_S=\{s_1,s_2,\ldots,s_{(k-3)n}\}$, $(m-n)$ \emph{dummy resources} $\Resources_D=\{d_1,\ldots,d_{m-n}\}$, and
					a \emph{special resource} $s^*$, i.e.,  $\Resources=\Resources_N\cup \Resources_S\cup \Resources_D\cup \{s^*\}$.
					\item For each cover agent $a_j$ and each normal resource $r_i$, the utility function is defined such that $u_j(r_i)=v$ if $x_i\in S_j$, and $u_j(r_i)=0$ otherwise.
					Besides, each cover agent values each small resource as $v$.
					Each cover agent values each dummy resource and the special resource $s^*$ as $kv$.
					Finally, the special agent $a^*$ values the special resource $s^*$ as $kv$ and values all other resources as 0.
				\end{custombul}

				We show that ${(X,C)}$ is a YES-instance if and only if $\mathcal{I}$ is a YES-instance.
				
				$(\Longrightarrow)$ Assume that ${(X,C)}$ is a YES-instance, then there is a subset
				$C'\subseteq C$ with $|C'|=n$ such that each $e \in X$ occurs in exactly one member of $C'$.
				For each $S_j \in C$, if $S_j\in C'$, then we allocate the 3 corresponding normal resources to $a_j$
				resulting in value that $a_j$ gets being exactly 3 for now.
				Then, $a_j$ will also get $k-3$ small resources, finally getting the
				value $u \,(=kv)$. If $S_j\notin C'$, then we allocate 1 dummy resource to $a_j$, which also
				results in value~$u$.
				In addition, the special agent will get the special resource which is valued at exactly $u$.
				It is easy to check that every agent gets utility $u=kv$ and values other agents' bundle by at most $u=kv$.
				Thus, $\mathcal{I}$ is also a YES-instance.
				
				$(\Longleftarrow)$ Assume that there is a solution for the constructed instance $\mathcal{I}$ of \efpashortGen{$\esw$}.
				Since in this solution each agent has to get a bundle with a positive value, the special agent will get the special resource~$s^*$.
				Then, each of the cover agents will require a bundle of value at least $u=kv$.
				Since the total value that all the $m$ cover agents can receive is at most $3nv+(k-3)nv+(m-n)kv=mkv$, the value that each cover agent receives should be exactly $kv$.
				Notice that $m-n$ dummy resources can be allocated to $m-n$ cover agents, so the remaining $n$ agents get all the normal and small resources.
				Since each remaining agent can receive at most value $3v$ from the normal resources, we conclude that each of them gets $3$ normal resources they like and $k-3$~small resources.
				Let $I_j=\{i_{ja}, i_{jb}, i_{jc}\}$ be the normal resources received by each remaining agent $a_j$.
				Then we can find $n$ corresponding sets $S_j=\{x_{ja}, x_{jb}, x_{jc}\}$ from $C$, which are pairwise disjoint.
				This induces a feasible solution $C'$ for ${(X,C)}$.
			\end{proof}
			
			Next we consider the case with $u=2v$. The distinctive feature of the
			following proof, 
			lies in our creation of
			standard agents and special resources as benchmarks, ensuring that the value
			of the bundle desired by each agent exceeds a certain constant value.
			Additionally, we introduce a large number of special ``observer'' agents and corresponding blank resources to monitor potential combinations of resources that may interfere with the reduction.
			
			\begin{lemma} \label{lem:ter-1-2}
				\efpashortGen{$\esw$} with ternary utility values $\{0,v,u\}$, $u=2v>0$, and $t=1$  is strongly NP-hard. 
			\end{lemma}

			\begin{proof}
				The hardness proof also proceeds by a reduction from X3C.
				Given an instance $(X,C)$ of the \textsc{X3C}, we construct an instance of $\mathcal{I} = (\Resources,\Agents,(u_a)_{a\in \Agents},t=1)$ of the \efpashortGen{$\esw$} problem as follows.
				\begin{custombul}
					\item There are $m$ \emph{cover agents} $\Agents_C=\{a_1, a_2, \ldots, a_m\}$, 3  \emph{standard agents} $b, c, d$, and a set~$\Observers$ of \emph{observers} (of finite size to be specified later), i.e., $\Agents=\Agents_C\cup \{b,c,d\}\cup \Observers$.
					\item There $3n$ \emph{normal resources} $\Resources_N=\{r_1,r_2,\ldots,r_{3n}\}$, $n$ \emph{small resources} $\Resources_S=\{s_1,s_2,\ldots,s_n\}$ , $2(m-n)$ \emph{dummy resources} $\Resources_D=\{d_1,\ldots,d_{2(m-n)}\}$
					and a finite number of \emph{blank resources} $\Resources_B$ (where
					$|\Resources_B|=2|\Observers|$) and 4 \emph{special resources}
					$r^*_1,r^*_2,r^*_3,r^*_4$, i.e.,  $\Resources=\Resources_N\cup \Resources_S\cup \Resources_D \cup \Resources_B\cup \{r^*_1,r^*_2,r^*_3,r^*_4\} $.
					\item For each cover agent $a_j$ and each normal resource $r_i$, the utility function is defined such that $u_j(r_i)=v$ if $x_i\in S_j$ and $u_j(r_i)=0$ otherwise.
					Besides, each cover agent values each small resource as $v$.
					In addition, each cover agent values each dummy resource and each special resource as $2v$.
					The cover agents are not interested in blank resources.
					\item For each standard agent and each special resource, the utility
					function is defined in~Table~\ref{tab:ter-1-2} and the standard
					agents are not interested in any of the other resources:
					{
						\renewcommand{\arraystretch}{1.25}
						\setlength{\tabcolsep}{12pt}
						\begin{table}[t]
							\caption{Agent's utility functions in the proof
								of~\Cref{lem:ter-1-2}.}
							\label{tab:ter-1-2}
							{
								\centering
								\begin{tabular}{c|ccc}
									& $b$ & $c$ & $d$ \\
									\hline
									$r^*_1$ & $2v$ & $2v$ & 0 \\
									$r^*_2$ & 0 & $v$ & 0 \\
									$r^*_3$ & $2v$ & $2v$ & $2v$ \\
									$r^*_4$ & 0 & $v$ & 0 \\
								\end{tabular}
								
							}
						\end{table}
					}


					\item Each observer assigns value~$2$ to each blank resource and each
					special resource.
					In particular, there are three different kinds of observers. Listing
					only resources for which the observers have a non-zero value, we define
					them as follows:
					(1)
					Each observer $w_{i,j;k}$ of type~1 values the two normal resources $r_i,r_j$, and one dummy resource $d_k$ at $2v$, respectively. 
					(2)
					Each observer $w'_{i;j;k}$ of type~2 values the normal resource $r_i$ and the dummy resource $d_j$ and the small resource $s_k$ at $2v$, respectively. 
					(3) An observer $w^*$ values every small resource and every dummy
					resource at~$2v$.
				\end{custombul}
				
				Overall, we create $ \tbinom{3n}{2}\cdot m+3m \cdot n \cdot 2(m-n)+1 $ observers.
				Thus, there are ${O}(m^2n)$ numbers of observers and blank resources.
				Assuming that there is a solution for the constructed instance
				$\mathcal{I}$ of \efpashortGen{$\esw$}, we have the following
				observations.
				\begin{enumerate}
					\item We first consider the standard agents.
					Since each agent has to get a positive value, standard agent $d$ will get $r^*_3$.
					Then, standard agent $b$ will get $r^*_1$ and standard agent $c$ will get $r^*_2$ and $r^*_4$.
					Since $c$ gets $r^*_2$ and $r^*_4$, each of the cover agents and the observers will require a value of at least $4v$.
					\item The normal resources, dummy resources and small resources can only be allocated to the $m$ cover agents.
					This is because the cover agents are not interested in the blank resources and the sum of the value that these three kinds of resources can provide is at most $4mv$.
					\item 
					Each cover agent gets utility exactly $4v$.
					Thus, if a cover agent gets 2 dummy resources, it cannot get any other resources.
					\item Since the observers can only get blank resources, each observer will get exactly two blank resources.
					\item From the previous Observations 1--4, we can claim that each resource is allocated to one agent in this allocation.
					\item No cover agent can get three different kinds of resources.
					Otherwise, some observer  $w'_{i;j;k}$ of type~2 will envy.
					\item No cover agent can get one dummy resource and one small resource.
					Otherwise, this agent needs another resource to ensure the bundle is of value at least~$4v$.
					However, this resource cannot be a normal resource according to Observation 6, and it cannot be a small or dummy resource since otherwise observer~$w^*$ would envy this agent.
					\item No cover agent receives one dummy resource and one normal resource.
					Otherwise, the agent needs another resource. Yet, neither can it be a
					dummy nor a normal resource 
					because of, respectively, the type 2 and~1 observers.
					\item It follows from Observations 3 and 6--8 above, that if
					some cover agent gets a dummy resource then it will get exactly two
					dummy resources and nothing else.
					Thus, there are $m-n$ cover agents who get $2(m-n)$ two dummy resources.
					\item The remaining $n$~cover agents get some normal resources and small
					resources and each of them gets exactly 1 small resource.
					This is because the value that each of them can get from the normal resources is at most $3v$.
					According to the pigeonhole principle, there is and can only be one small resource for each cover agent.
				\end{enumerate}
				
				We show that ${(X,C)}$ is a YES-instance if and only if $\mathcal{I}$ is a YES-instance.
				
				$(\Longrightarrow)$ Since ${(X,C)}$ is a YES-instance, there is a subset
				$C'\subseteq C$ with $|C'|=n$ such that every element of $X$ occurs in
				exactly one member of $C'$.  If $S_j\in C'$, we allocate the 3
				corresponding normal resources to each $a_j$ such that the value that
				$a_j$ can get is exactly $3v$ for now.  In addition, each~$a_j$ will
				also get 1 small resource and finally get the value $4v$.  If $S_j\notin
				C'$, we allocate 2 dummy resources and the value is also $4v$. Further,
				each observer gets 2 blank resources and the value is also $4v$.
				Finally, $b$ gets $r^*_1$, $c$ gets $r^*_2$ and $r^*_4$, $d$ gets
				$r^*_3$. In this case, no agent is envious. Thus, $\mathcal{I}$ is also a
				YES-instance.
				
				$(\Longleftarrow)$ Since $\mathcal{I}$ is a YES-instance, combining the observations above,
				note that there are $n$ agents $a_j$ who only get three normal resources
				$I_j=\{i_{ja}, i_{jb}, i_{jc}\}$ and one small resource such that we can
				find $n$ corresponding sets $S_j=\{x_{ja}, x_{jb}, x_{jc}\}$.
				We can find exactly $n$ such disjoint sets, which induces a feasible solution $C'$. Thus, ${(X,C)}$ is a YES-instance.
			\end{proof}

			Finally, we consider the case when $u$ is not divisible by $v$. The following proof, while similar to the previous one, involves additional considerations. These arise primarily because $u$ may be significantly greater than $v$. For some agents, in order to achieve a value exceeding $u$ or even $2u$ solely through resources valued at $v$, they would need to acquire a multiple of these resources.
			
			\begin{lemma} \label{lem:ter-v-kv+c}
				With ternary utility values $\{0,v,u=kv+c\}$ for $v>c>0,k>0$ \efpashortGen{$\esw$} is strongly NP-hard and $t=1$.
			\end{lemma}
		
			{
				
				\begin{proof}
					This final case builds up on ideas from Case 2, but has some more technicalities
					and uses more involved auxiliary agents and resources.
					The hardness proof is again realized via a reduction from X3C.
					Given an instance $(X,C)$ of the \textsc{X3C} problem, we construct an instance $\mathcal{I} = (\Resources,\Agents,(u_a)_{a\in \Agents},t=1)$ of the \efpashortGen{$\esw$} problem as follows.
					
					We have the following resources.
					\begin{custombul}
						\item A set of \emph{element resources} $\Resources_X=\{r_1,r_2,\ldots,r_{3n}\}$ as well as
						a set of \emph{dummy resources} $\Resources_D=\{d_1,\ldots,d_{3(m-n)}\}$.
						\item Three sets of \emph{special resources}: namely \emph{booster resources}~$\Resources_B$ as well as
						\emph{guard resources} $\Resources^{\text{rrd}}_G$ and $\Resources^{\text{rdd}}_G$,
						whose cardinalities are specified later.
						\item Altogether, $\Resources=\Resources_X\cup \Resources_D \cup \Resources^B \cup \Resources^{\text{rrd}}_G \cup \Resources^{\text{rdd}}_G$.
					\end{custombul}
					
					We have the following agents.
					\begin{custombul}
						\item A set of $m$~\emph{cover agents} $\Agents_C=\{a_1, a_2, \ldots, a_m\}$.
						\item A set of three~\emph{(utility) booster agents} $\Agents_B=\{b_1, b_2, b_3\}$.
						\item A set of $\tbinom{3n}{2}\cdot 3(m-n)$~\emph{$\text{rrd}$-guards} $\Agents_G^{\text{rrd}}=\{g^{\text{rrd}}(r,r',d) \mid
						r,r' \in \mathcal{R}_X, d \in \mathcal{R}_D, r \neq r'\}$.
						\item A set of $\tbinom{3(m-n)}{2}\cdot 3n$~\emph{$\text{rdd}$-guards} $\Agents_G^{\text{rdd}}=\{g^{\text{rrd}}(r,d,d') \mid
						r \in \mathcal{R}_X, d,d' \in \mathcal{R}_D, d \neq d'\}$.
					\end{custombul}
					
					Before we go into the formal proof, we provide some intuition:
					\begin{custombul}
						\item The cover agents together with the element and dummy resources are meant to encode the X3C solution:
						A cover agent corresponding to a set selected in the X3C solution gets a bundle of three element resources and a cover agent corresponding to a set that is not selected gets a bundle of three dummy resources.
						\item Booster agents are used to boost the minimum utility value of the other agents' bundle:
						They need to obtain predetermined bundles of booster resources and the other agents
						also have some value for some of these resources.
						To avoid envy towards the booster agents, the bundles of the other agents need to have
						a specific value for them.
						\item Guard agents ensure that cover agents can only get bundles that either contain three element
						or three dummy resources.
						Each guard ``forbids'' a specific combination mixed bundles.
					\end{custombul}
					
					Next, we fully specify the special resources.
					To do so, we define that $k_1=k+1$ and $k_2=\min\{k'\in \mathbb{N} \mid k'*v>2u\}$.
					Now, $\Resources_B=\Resources_B^{b_1}\cup\Resources_B^{b_2}\cup\Resources_B^{b_3}$ with
					$\Resources_B^{b_1}=\{s^{b_1}\}$, $\Resources_B^{b_2}=\{s^{b_2}_1,s^{b_2}_2,\dots,s^{b_2}_{k_1}\}$,
					and $\Resources_B^{b_3}=\{s^{b_3}_1,s^{b_3}_2,\dots,s^{b_3}_{k_2}\}$.
					Moreover, the guards resources are
					$\Resources^{\text{rrd}}_G=\{s^{\text{rrd}}_1,s^{\text{rrd}}_2,\dots,s^{\text{rrd}}_{k_2\cdot|\Agents_G^{\text{rrd}}|}\}$
					and $\Resources^{\text{rdd}}_G=\{s^{\text{rdd}}_1,s^{\text{rdd}}_2,\dots,s^{\text{rdd}}_{k_2\cdot|\Agents_G^{\text{rdd}}|}\}$.
					Note that $v\leq k v<u<k_1 v<2u<k_2 v<3u$.

					We now define the \emph{utility functions} of the agents by specifying non-zero utility values
					(that is, in all combinations not specified here, the agent assigns utility zero to the resource).
					
					\begin{custombul}
						\item For each cover agent~$a_j$ and each element resource $r_i$, we have $u_{a_j}(r_i)=u$
						if $x_i\in S_j$ (and $u_{a_j}(r_i)=0$ otherwise).
						Moreover, cover agents assign utility~$u$ to each dummy resource and to
						$s^{b_3}_1$, $s^{b_3}_2$, and $s^{b_3}_3$.
						\item The booster agent~$b_1$ assigns utility~$u$ to~$s^{b_1}$ only.
						\item The booster agent~$b_2$ assigns utility~$u$ to~$s^{b_1}$ and
						utility~$v$ to each resource from~$\Resources_B^{b_2}$.
						\item The booster agent~$b_3$ assigns utility~$u$ to~$s^{b_2}_1$ and~$s^{b_2}_2$, and
						utility~$v$ to each resource from~$\Resources_B^{b_3}$.
						\item The $\text{rrd}$-guard~$g^{\text{rrd}}(r,r',d)$ assigns utility~$u$
						to the element resources~$r$ and~$r'$ and to the dummy resource~$d$
						and utility~$v$ to each of the resources from~$\Resources^{\text{rrd}}_G$.
						He also assigns utility~$u$ to $s^{b_3}_1$ and $s^{b_3}_2$.
						\item The $\text{rdd}$-guard~$g^{\text{rrd}}(r,d,d')$ assigns utility~$u$
						to the element resource~$r$ and to the dummy resources~$d$ and~$d'$
						and utility~$v$ to each of the resources from~$\Resources^{\text{rdd}}_G$.
						He also assigns utility~$u$ to $s^{b_3}_1$ and $s^{b_3}_2$.
					\end{custombul}
					
					Finally, there are overall
					$m+3+\tbinom{3n}{2}\cdot 3(m-n) + \tbinom{3(m-n)}{2}\cdot 3n$ agents and
					$3n+3(m-n)
					+1+k_1+k_2
					+k_2\cdot\tbinom{3n}{2}\cdot 3(m-n)
					+k_2\cdot\tbinom{3(m-n)}{2}\cdot 3n$ resources.
					The the construction can definitely be performed in polynomial time.
					
					Assuming that there is a solution for the constructed instance $\mathcal{I}$, we have the following observations.
					\begin{enumerate}[label=Ob.\,\arabic*{}., ref=\arabic*, leftmargin=8ex]
						\item The agent $b_1$ will get $s^{b_1}$, otherwise the value of his bundle will be 0.
						\label[obs]{lem:ter-v-kv+c:obb1}
						\item The agent $b_2$ will get all resources from $\Resources_B^{b_2}$.
						Otherwise, he will envy $b_1$ who gets value of $u$ in his eyes.
						\label[obs]{lem:ter-v-kv+c:obb2}
						\item The agent $b_3$ will get all resources of $\Resources_B^{b_3}$.
						Otherwise, he will envy $b_2$ who gets value of $2u$ in his eyes.
						\label[obs]{lem:ter-v-kv+c:obb3}
						\item Each of the cover agents will get exactly three resources from $\Resources_X \cup \Resources_D$
						and get value of $3u$.
						Otherwise, some of them will envy booster agent~$b_3$ who gets value of~$3u$ in their eyes.
						By the pigeonhole principle, this implies that other agents (who are not cover agents) cannot get
						any element or dummy resource.
						\label[obs]{lem:ter-v-kv+c:obcover1}
						\item Each $\text{rrd}$-guard gets $k_2$~arbitrary resources from~$\Resources^{\text{rrd}}_G$
						and each $\text{rdd}$-guard gets $k_2$~arbitrary resources from~$\Resources^{\text{rdd}}_G$.
						If any of the guard agent gets fewer than $k_2$~of the corresponding resources,
						then the value of his bundle would be smaller than~$2u$ and he would envy booster agent~$b_3$.
						By pigeonhole principle, no guard agent can get more than~$k_2$ of the corresponding resources.
						\label[obs]{lem:ter-v-kv+c:obguard}
						\item It is not possible for any cover agent to get a bundle with resources from both~$R_X$ and~$R_D$.
						Assume that such agent $a'$~exists and gets~$\{r,r',d\}$ with~$r\neq r'\in\Resources_X$ and~$d\in\Resources_D$.
						Then, the $\text{rrd}$-guard~$g(r,r',d)$ would require at least $k_2+1$~resources
						from~$\Resources^{\text{rrd}}_G$ (to not envy~$a'$); a contradiction to the previous observation.
						Analogously (replacing rrd-guards by rdd-guards), no cover agent can
						get~$\{r,d,d'\}$ with~$r\in\Resources_X$ and~$d\neq d'\in\Resources_D$.
						\label[obs]{lem:ter-v-kv+c:obcover2}
					\end{enumerate}
					
					We show that ${(X,C)}$ is a YES-instance if and only if $\mathcal{I}$ is a YES-instance.
					
					$(\Longrightarrow)$ Assume that ${(X,C)}$ is a YES-instance, then there exists a subset $C'\subseteq C$ with $|C'|=n$ such that every element of $X$ occurs in exactly one member of $C'$.
					We construct a solution for $\mathcal{I}$ as follows.
					Booster agents will get resources as discussed in \Cref{lem:ter-v-kv+c:obb1,lem:ter-v-kv+c:obb2,lem:ter-v-kv+c:obb3} and guard-agents will get get resources as discussed in \Cref{lem:ter-v-kv+c:obguard}.
					For each $S_j \in C$, if $S_j\in C'$, we allocate the three corresponding element resources to~$a_j$.
					If $S_j\notin C'$, we allocate three arbitrary dummy resources to~$a_j$.
					Since there are no combinations of element resources and dummy resources (\cref{lem:ter-v-kv+c:obcover2}), no guard-agent will envy (\cref{lem:ter-v-kv+c:obcover1}).
					Therefore, no agent will envy in this case.
					Thus, $\mathcal{I}$ is also a YES-instance.
					
					$(\Longleftarrow)$ Assume that $\mathcal{I}$ is a YES-instance.
					According to the above observations,
					there is a set~$A^*$ of~$n$ cover agents who each get a bundle of three element resources.
					Let $A^*=\{a_{j_1},a_{j_2},\dots,a_{j_n}\}$.
					Now, observe that $C^*=\{S_{j_1},S_{j_2},\dots,S_{j_n}\}$ clearly form a feasible solution for~${(X,C)}$:
					Each element is covered exactly once since each element resource is assigned to exactly one of these
					cover agents. Moreover, the bundles of~$A^*$ correspond exactly to the item subsets from~$C^*$.
					Thus, ${(X,C)}$ is a YES-instance.
			\end{proof}
		}
		The claim of \Cref{thm:ter} follows from
		\Cref{lemma:ter+usw+mcar+size,lem:ter-1-3,lem:ter-1-2,lem:ter-v-kv+c}.

		\section{Conclusions}
		
		We studied how to allocate indivisible resources to agents in an envy-free
		manner by relaxing the common requirement that all resources must be
		allocated.
		We considered envy-free partial allocations that provide at least some
		utility or allocate some resources from both systematic or individual
		perspectives, and we obtained comprehensive results under various classes of
		utilities.
		While most of the problems we considered are generally NP-hard, we
		identified several tractable results for binary utilities by establishing interesting
		connections to matching problems on bipartite graphs.
		Notably, our tractable results imply that, at least for binary utilities, if
		the goal is to allocate some resources or provide some utility to agents, then
		the problem of finding envy-free partial allocations (or confirming their
		non-existence) can be efficiently solved.
		Complementing the well-known NP-hardness of finding envy-free complete
		allocations, our results provide a more fine-grained understanding of the
		computational complexity of finding efficient envy-free allocations.
		
		Our work can be extended in several directions.
		First, we show a stark contrast: some cases are tractable under binary
		utilities but all scenarios become NP-hard under ternary utilities. It is
		worth further exploring this frontier, in particular, bivalued utilities other
		than the combination of $0$ and~$1$, that lie between binary and ternary
		utilities.
		%
		%
		In the supplementary material, we provide some initial results for $1/2$ utilities. 
		When $t=1$ all the four efficiency measures are equivalent, and we can reduce
		the problem to the case where each agent can get at most two resources. 
		Second, we assumed all resources are goods. 
		A natural extension is to study chores or mixed resources.
		For chores, the case of a planner who wants to distribute as many tasks to
		agents as possible well justifies our measures $\size$ and $\mcar$.
		Here a relevant result is that for chores and binary values (or even binary
		marginals), there always exists an envy-free allocation with at most $n-1$
		unallocated resources~\cite{DBLP:conf/faw/TaoWYZ24}.
		Finally, applying our setting for alternative fairness notions, such as
		equitability, instead of envy-freeness offers another research direction.
		We note that for identical utilities,
		these two fairness notions are equivalent.

		
		
		\section*{Acknowledgement}
			
			This project has received funding from the European Research Council (ERC)
			under the European Union’s Horizon 2020 research and innovation programme
			(grant agreement No 101002854). JL acknowledges support from the National Natural Science Foundation of China (No. 12301412).
			
			{\centering
				\includegraphics[width=2cm]{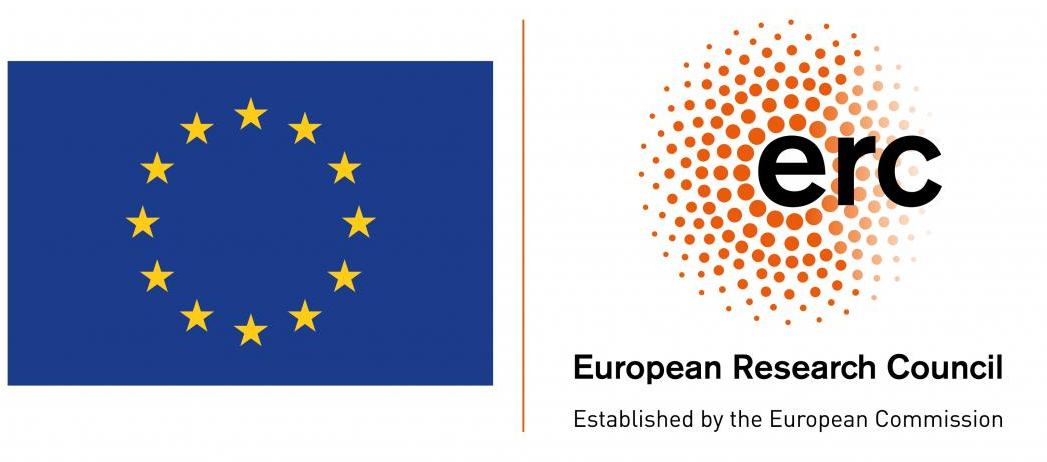}
				
			}

		\bibliographystyle{ACM-Reference-Format}
		\balance 
		\bibliography{refnew}


\begin{thebibliography}{27}


\ifx \showCODEN    \undefined \def \showCODEN     #1{\unskip}     \fi
\ifx \showDOI      \undefined \def \showDOI       #1{#1}\fi
\ifx \showISBNx    \undefined \def \showISBNx     #1{\unskip}     \fi
\ifx \showISBNxiii \undefined \def \showISBNxiii  #1{\unskip}     \fi
\ifx \showISSN     \undefined \def \showISSN      #1{\unskip}     \fi
\ifx \showLCCN     \undefined \def \showLCCN      #1{\unskip}     \fi
\ifx \shownote     \undefined \def \shownote      #1{#1}          \fi
\ifx \showarticletitle \undefined \def \showarticletitle #1{#1}   \fi
\ifx \showURL      \undefined \def \showURL       {\relax}        \fi
\providecommand\bibfield[2]{#2}
\providecommand\bibinfo[2]{#2}
\providecommand\natexlab[1]{#1}
\providecommand\showeprint[2][]{arXiv:#2}

\bibitem[Aigner-Horev and Segal-Halevi(2022)]%
        {aigner2022}
\bibfield{author}{\bibinfo{person}{Elad Aigner-Horev} {and}
  \bibinfo{person}{Erel Segal-Halevi}.} \bibinfo{year}{2022}\natexlab{}.
\newblock \showarticletitle{Envy-free matchings in bipartite graphs and their
  applications to fair division}.
\newblock \bibinfo{journal}{\emph{Information Sciences}}  \bibinfo{volume}{587}
  (\bibinfo{year}{2022}), \bibinfo{pages}{164--187}.
\newblock


\bibitem[Aziz et~al\mbox{.}(2015)]%
        {AzizGMW15}
\bibfield{author}{\bibinfo{person}{Haris Aziz}, \bibinfo{person}{Serge
  Gaspers}, \bibinfo{person}{Simon Mackenzie}, {and} \bibinfo{person}{Toby
  Walsh}.} \bibinfo{year}{2015}\natexlab{}.
\newblock \showarticletitle{Fair assignment of indivisible objects under
  ordinal preferences}.
\newblock \bibinfo{journal}{\emph{Artificial Intelligence}}
  \bibinfo{volume}{227} (\bibinfo{year}{2015}), \bibinfo{pages}{71--92}.
\newblock


\bibitem[Aziz et~al\mbox{.}(2023)]%
        {AzizHMS23}
\bibfield{author}{\bibinfo{person}{Haris Aziz}, \bibinfo{person}{Xin Huang},
  \bibinfo{person}{Nicholas Mattei}, {and} \bibinfo{person}{Erel
  Segal-Halevi}.} \bibinfo{year}{2023}\natexlab{}.
\newblock \showarticletitle{Computing welfare-Maximizing fair allocations of
  indivisible goods}.
\newblock \bibinfo{journal}{\emph{European Journal of Operational Research}}
  \bibinfo{volume}{307}, \bibinfo{number}{2} (\bibinfo{year}{2023}),
  \bibinfo{pages}{773--784}.
\newblock


\bibitem[Aziz et~al\mbox{.}(2016)]%
        {AzizSW16}
\bibfield{author}{\bibinfo{person}{Haris Aziz}, \bibinfo{person}{Ildik{\'{o}}
  Schlotter}, {and} \bibinfo{person}{Toby Walsh}.}
  \bibinfo{year}{2016}\natexlab{}.
\newblock \showarticletitle{Control of Fair Division}. In
  \bibinfo{booktitle}{\emph{Proceedings of the 25th International Joint
  Conference on Artificial Intelligence ({IJCAI} '16)}}.
  \bibinfo{publisher}{{IJCAI/AAAI} Press}, \bibinfo{pages}{67--73}.
\newblock


\bibitem[Babaioff et~al\mbox{.}(2021)]%
        {BabaioffEF21}
\bibfield{author}{\bibinfo{person}{Moshe Babaioff}, \bibinfo{person}{Tomer
  Ezra}, {and} \bibinfo{person}{Uriel Feige}.} \bibinfo{year}{2021}\natexlab{}.
\newblock \showarticletitle{Fair and Truthful Mechanisms for Dichotomous
  Valuations}. In \bibinfo{booktitle}{\emph{Proceedings of the 35th AAAI
  Conference on Artificial Intelligence ({AAAI} '21)}}.
  \bibinfo{publisher}{AAAI Press}, \bibinfo{pages}{5119--5126}.
\newblock


\bibitem[Barman et~al\mbox{.}(2018)]%
        {BarmanKV18}
\bibfield{author}{\bibinfo{person}{Siddharth Barman},
  \bibinfo{person}{Sanath~Kumar Krishnamurthy}, {and} \bibinfo{person}{Rohit
  Vaish}.} \bibinfo{year}{2018}\natexlab{}.
\newblock \showarticletitle{Greedy Algorithms for Maximizing Nash Social
  Welfare}. In \bibinfo{booktitle}{\emph{Proceedings of the 17th International
  Conference on Autonomous Agents and MultiAgent Systems, ({AAMAS} '18)}}.
  \bibinfo{publisher}{International Foundation for Autonomous Agents and
  Multiagent Systems Richland, SC, {USA} / {ACM}}, \bibinfo{pages}{7--13}.
\newblock


\bibitem[Berger et~al\mbox{.}(2022)]%
        {BergerCFF22}
\bibfield{author}{\bibinfo{person}{Ben Berger}, \bibinfo{person}{Avi Cohen},
  \bibinfo{person}{Michal Feldman}, {and} \bibinfo{person}{Amos Fiat}.}
  \bibinfo{year}{2022}\natexlab{}.
\newblock \showarticletitle{Almost Full EFX Exists for Four Agents}. In
  \bibinfo{booktitle}{\emph{Proceedings of the 26th AAAI Conference on
  Artificial Intelligence ({AAAI} '22)}}. \bibinfo{publisher}{{AAAI} Press},
  \bibinfo{pages}{4826--4833}.
\newblock


\bibitem[Beynier et~al\mbox{.}(2019)]%
        {BeynierCGHLMW19}
\bibfield{author}{\bibinfo{person}{Aur{\'{e}}lie Beynier},
  \bibinfo{person}{Yann Chevaleyre}, \bibinfo{person}{Laurent Gourv{\`{e}}s},
  \bibinfo{person}{Ararat Harutyunyan}, \bibinfo{person}{Julien Lesca},
  \bibinfo{person}{Nicolas Maudet}, {and} \bibinfo{person}{Ana{\"{e}}lle
  Wilczynski}.} \bibinfo{year}{2019}\natexlab{}.
\newblock \showarticletitle{Local envy-freeness in house allocation problems}.
\newblock \bibinfo{journal}{\emph{Autonomous Agents and Multi-Agent Systems}}
  \bibinfo{volume}{33}, \bibinfo{number}{5} (\bibinfo{year}{2019}),
  \bibinfo{pages}{591--627}.
\newblock


\bibitem[Bliem et~al\mbox{.}(2016)]%
        {DBLP:conf/ijcai/BliemBN16}
\bibfield{author}{\bibinfo{person}{Bernhard Bliem}, \bibinfo{person}{Robert
  Bredereck}, {and} \bibinfo{person}{Rolf Niedermeier}.}
  \bibinfo{year}{2016}\natexlab{}.
\newblock \showarticletitle{Complexity of Efficient and Envy-Free Resource
  Allocation: Few Agents, Resources, or Utility Levels}. In
  \bibinfo{booktitle}{\emph{Proceedings of the 25th International Joint
  Conference on Artificial Intelligence, ({IJCAI} '16)}}.
  \bibinfo{publisher}{{IJCAI/AAAI} Press}, \bibinfo{pages}{102--108}.
\newblock


\bibitem[Boehmer et~al\mbox{.}(2024)]%
        {DBLP:journals/aamas/BoehmerBHKL24}
\bibfield{author}{\bibinfo{person}{Niclas Boehmer}, \bibinfo{person}{Robert
  Bredereck}, \bibinfo{person}{Klaus Heeger}, \bibinfo{person}{Dusan Knop},
  {and} \bibinfo{person}{Junjie Luo}.} \bibinfo{year}{2024}\natexlab{}.
\newblock \showarticletitle{Multivariate algorithmics for eliminating envy by
  donating goods}.
\newblock \bibinfo{journal}{\emph{Autonomous Agents and Multi-Agent Systems}}
  \bibinfo{volume}{38}, \bibinfo{number}{2} (\bibinfo{year}{2024}),
  \bibinfo{pages}{43}.
\newblock


\bibitem[Bouveret et~al\mbox{.}(2016)]%
        {BCM16}
\bibfield{author}{\bibinfo{person}{Sylvain Bouveret}, \bibinfo{person}{Yann
  Chevaleyre}, {and} \bibinfo{person}{Nicolas Maudet}.}
  \bibinfo{year}{2016}\natexlab{}.
\newblock \showarticletitle{Fair Allocation of Indivisible Goods}.
\newblock In \bibinfo{booktitle}{\emph{Handbook of Computational Social
  Choice}}, \bibfield{editor}{\bibinfo{person}{F.~Brandt},
  \bibinfo{person}{V.~Conitzer}, \bibinfo{person}{U.~Endriss},
  \bibinfo{person}{J.~Lang}, {and} \bibinfo{person}{A.~D. Procaccia}} (Eds.).
  \bibinfo{publisher}{Cambridge University Press}, Chapter~12.
\newblock


\bibitem[Bouveret and Lang(2008)]%
        {BL08}
\bibfield{author}{\bibinfo{person}{Sylvain Bouveret} {and}
  \bibinfo{person}{J{\'e}r\^{o}me Lang}.} \bibinfo{year}{2008}\natexlab{}.
\newblock \showarticletitle{Efficiency and Envy-freeness in Fair Division of
  Indivisible Goods: Logical Representation and Complexity}.
\newblock \bibinfo{journal}{\emph{Journal of Artificial Intelligence Research}}
  \bibinfo{volume}{32}, \bibinfo{number}{1} (\bibinfo{year}{2008}),
  \bibinfo{pages}{525--564}.
\newblock


\bibitem[Bu et~al\mbox{.}(2022)]%
        {Bu22}
\bibfield{author}{\bibinfo{person}{Xiaolin Bu}, \bibinfo{person}{Zihao Li},
  \bibinfo{person}{Shengxin Liu}, \bibinfo{person}{Jiaxin Song}, {and}
  \bibinfo{person}{Biaoshuai Tao}.} \bibinfo{year}{2022}\natexlab{}.
\newblock \showarticletitle{On the Complexity of Maximizing Social Welfare
  within Fair Allocations of Indivisible Goods}.
\newblock \bibinfo{journal}{\emph{CoRR}}  \bibinfo{volume}{abs/2205.14296}
  (\bibinfo{year}{2022}).
\newblock


\bibitem[Budish(2011)]%
        {Bud11}
\bibfield{author}{\bibinfo{person}{Eric Budish}.}
  \bibinfo{year}{2011}\natexlab{}.
\newblock \showarticletitle{The Combinatorial Assignment Problem: Approximate
  Competitive Equilibrium from Equal Incomes}.
\newblock \bibinfo{journal}{\emph{Journal of Political Economy}}
  \bibinfo{volume}{119}, \bibinfo{number}{6} (\bibinfo{year}{2011}),
  \bibinfo{pages}{1061--1103}.
\newblock


\bibitem[Caragiannis et~al\mbox{.}(2019a)]%
        {DBLP:conf/ec/CaragiannisGH19}
\bibfield{author}{\bibinfo{person}{Ioannis Caragiannis}, \bibinfo{person}{Nick
  Gravin}, {and} \bibinfo{person}{Xin Huang}.}
  \bibinfo{year}{2019}\natexlab{a}.
\newblock \showarticletitle{Envy-Freeness Up to Any Item with High {N}ash
  Welfare: The Virtue of Donating Items}. In
  \bibinfo{booktitle}{\emph{Proceedings of the 20th ACM Conference on Economics
  and Computation ({EC} '19)}}. \bibinfo{publisher}{{ACM}},
  \bibinfo{pages}{527--545}.
\newblock


\bibitem[Caragiannis et~al\mbox{.}(2019b)]%
        {CKMPSW19}
\bibfield{author}{\bibinfo{person}{Ioannis Caragiannis}, \bibinfo{person}{David
  Kurokawa}, \bibinfo{person}{Herv{\'e} Moulin}, \bibinfo{person}{Ariel~D.
  Procaccia}, \bibinfo{person}{Nisarg Shah}, {and} \bibinfo{person}{Junxing
  Wang}.} \bibinfo{year}{2019}\natexlab{b}.
\newblock \showarticletitle{The Unreasonable Fairness of Maximum Nash Welfare}.
\newblock \bibinfo{journal}{\emph{ACM Transactions on Economics and
  Computation}} \bibinfo{volume}{7}, \bibinfo{number}{3}
  (\bibinfo{year}{2019}), \bibinfo{pages}{12:1--12:32}.
\newblock


\bibitem[Chaudhury et~al\mbox{.}(2021)]%
        {DBLP:journals/siamcomp/ChaudhuryKMS21}
\bibfield{author}{\bibinfo{person}{Bhaskar~Ray Chaudhury},
  \bibinfo{person}{Telikepalli Kavitha}, \bibinfo{person}{Kurt Mehlhorn}, {and}
  \bibinfo{person}{Alkmini Sgouritsa}.} \bibinfo{year}{2021}\natexlab{}.
\newblock \showarticletitle{A Little Charity Guarantees Almost Envy-Freeness}.
\newblock \bibinfo{journal}{\emph{SIAM J. Comput.}} \bibinfo{volume}{50},
  \bibinfo{number}{4} (\bibinfo{year}{2021}), \bibinfo{pages}{1336--1358}.
\newblock


\bibitem[Dorn et~al\mbox{.}(2021)]%
        {DBLP:journals/algorithmica/DornHS21}
\bibfield{author}{\bibinfo{person}{Britta Dorn}, \bibinfo{person}{Ronald de
  Haan}, {and} \bibinfo{person}{Ildik{\'{o}} Schlotter}.}
  \bibinfo{year}{2021}\natexlab{}.
\newblock \showarticletitle{Obtaining a Proportional Allocation by Deleting
  Items}.
\newblock \bibinfo{journal}{\emph{Algorithmica}} \bibinfo{volume}{83},
  \bibinfo{number}{5} (\bibinfo{year}{2021}), \bibinfo{pages}{1559--1603}.
\newblock


\bibitem[Fitzsimmons et~al\mbox{.}(2024)]%
        {DBLP:journals/corr/abs-2403-00943}
\bibfield{author}{\bibinfo{person}{Zack Fitzsimmons}, \bibinfo{person}{Vignesh
  Viswanathan}, {and} \bibinfo{person}{Yair Zick}.}
  \bibinfo{year}{2024}\natexlab{}.
\newblock \showarticletitle{On the Hardness of Fair Allocation under Ternary
  Valuations}.
\newblock \bibinfo{journal}{\emph{CoRR}}  \bibinfo{volume}{abs/2403.00943}
  (\bibinfo{year}{2024}).
\newblock


\bibitem[Gan et~al\mbox{.}(2019)]%
        {GanSV19}
\bibfield{author}{\bibinfo{person}{Jiarui Gan}, \bibinfo{person}{Warut
  Suksompong}, {and} \bibinfo{person}{Alexandros~A. Voudouris}.}
  \bibinfo{year}{2019}\natexlab{}.
\newblock \showarticletitle{Envy-freeness in house allocation problems}.
\newblock \bibinfo{journal}{\emph{Mathematical Social Sciences}}
  \bibinfo{volume}{101} (\bibinfo{year}{2019}), \bibinfo{pages}{104--106}.
\newblock


\bibitem[Garey and Johnson(1979)]%
        {DBLP:books/fm/GareyJ79}
\bibfield{author}{\bibinfo{person}{M.~R. Garey} {and} \bibinfo{person}{David~S.
  Johnson}.} \bibinfo{year}{1979}\natexlab{}.
\newblock \bibinfo{booktitle}{\emph{Computers and Intractability: {A} Guide to
  the Theory of NP-Completeness}}.
\newblock \bibinfo{publisher}{W. H. Freeman}.
\newblock


\bibitem[Halpern et~al\mbox{.}(2020)]%
        {DBLP:conf/wine/0002PP020}
\bibfield{author}{\bibinfo{person}{Daniel Halpern}, \bibinfo{person}{Ariel~D.
  Procaccia}, \bibinfo{person}{Alexandros Psomas}, {and}
  \bibinfo{person}{Nisarg Shah}.} \bibinfo{year}{2020}\natexlab{}.
\newblock \showarticletitle{Fair Division with Binary Valuations: One Rule to
  Rule Them All}. In \bibinfo{booktitle}{\emph{Proceedings of 16th
  International Conference on Web and Internet Economics ({WINE} '20)}}
  \emph{(\bibinfo{series}{Lecture Notes in Computer Science},
  Vol.~\bibinfo{volume}{12495})}. \bibinfo{publisher}{Springer},
  \bibinfo{pages}{370--383}.
\newblock


\bibitem[Hosseini et~al\mbox{.}(2020)]%
        {HosseiniSVWX20}
\bibfield{author}{\bibinfo{person}{Hadi Hosseini}, \bibinfo{person}{Sujoy
  Sikdar}, \bibinfo{person}{Rohit Vaish}, \bibinfo{person}{Hejun Wang}, {and}
  \bibinfo{person}{Lirong Xia}.} \bibinfo{year}{2020}\natexlab{}.
\newblock \showarticletitle{Fair Division Through Information Withholding}. In
  \bibinfo{booktitle}{\emph{The 24th AAAI Conference on Artificial Intelligence
  ({AAAI} '20)}}. \bibinfo{publisher}{{AAAI} Press},
  \bibinfo{pages}{2014--2021}.
\newblock


\bibitem[Kamiyama et~al\mbox{.}(2021)]%
        {KamiyamaMS21}
\bibfield{author}{\bibinfo{person}{Naoyuki Kamiyama}, \bibinfo{person}{Pasin
  Manurangsi}, {and} \bibinfo{person}{Warut Suksompong}.}
  \bibinfo{year}{2021}\natexlab{}.
\newblock \showarticletitle{On the complexity of fair house allocation}.
\newblock \bibinfo{journal}{\emph{Operations Research Letters}}
  \bibinfo{volume}{49}, \bibinfo{number}{4} (\bibinfo{year}{2021}),
  \bibinfo{pages}{572--577}.
\newblock


\bibitem[Ramshaw and Tarjan(2012)]%
        {ramshaw2012minimum}
\bibfield{author}{\bibinfo{person}{Lyle Ramshaw} {and}
  \bibinfo{person}{Robert~E Tarjan}.} \bibinfo{year}{2012}\natexlab{}.
\newblock \showarticletitle{On minimum-cost assignments in unbalanced bipartite
  graphs}.
\newblock \bibinfo{journal}{\emph{technical report}}  \bibinfo{volume}{20}
  (\bibinfo{year}{2012}).
\newblock


\bibitem[Tao et~al\mbox{.}(2024)]%
        {DBLP:conf/faw/TaoWYZ24}
\bibfield{author}{\bibinfo{person}{Biaoshuai Tao}, \bibinfo{person}{Xiaowei
  Wu}, \bibinfo{person}{Ziqi Yu}, {and} \bibinfo{person}{Shengwei Zhou}.}
  \bibinfo{year}{2024}\natexlab{}.
\newblock \showarticletitle{On the Existence of {EFX} (and Pareto-Optimal)
  Allocations for Binary Chores}. In \bibinfo{booktitle}{\emph{In Procedings of
  the 18th International Joint Conference on Frontiers of Algorithmics
  ({IJTCS-FAW} '24)}} \emph{(\bibinfo{series}{Lecture Notes in Computer
  Science}, Vol.~\bibinfo{volume}{14752})}. \bibinfo{publisher}{Springer},
  \bibinfo{pages}{33--52}.
\newblock


\bibitem[Walsh(2015)]%
        {Wal15}
\bibfield{author}{\bibinfo{person}{Toby Walsh}.}
  \bibinfo{year}{2015}\natexlab{}.
\newblock \showarticletitle{Challenges in Resource and Cost Allocation}. In
  \bibinfo{booktitle}{\emph{Proceedings of the 29th AAAI Conference on
  Artificial Intelligence ({AAAI} '15)}}. \bibinfo{publisher}{{AAAI} Press},
  \bibinfo{pages}{4073--4077}.
\newblock


\end{thebibliography}
		\clearpage
		\appendix
		\section*{Supplementary Material}

		\section{Bivalued valuations}\label{app:bivalued}

		When $u_i(a_j) \in \{1,2\}$ for any agent $a_i \in \Agents$ and any resource $r_j \in \Resources$, the four efficiency measures are equivalent. 
		
		\begin{lemma}\label{lem:bivalues-equivalent}
			When $t=1$, for any~$\Efficiency_1,\Efficiency_2  \in \{\usw, \esw, \size, \mcar\}$, \efpashortGen{$\Efficiency_1$} and \efpashortGen{$\Efficiency_2$} are equivalent.
		\end{lemma}
		
		\begin{proof}
			Since $u_i(a_j) \in \{1,2\}$, if an envy-free allocation~$\allocation$ satisfies that $\usw(\allocation)\geq1$, then it follows that $\size(\allocation)\geq1$.
			By envy-freeness, $\size(\allocation)\geq1$ implies that $\mcar(\allocation)\geq1$.
			By positive value of every resource, $\mcar(\allocation)\geq1$ implies that $\esw(\allocation)\geq1$, which implies that $\usw(\allocation)\geq1$.
		\end{proof}
		
		We show that we can reduce the problem to the case where each agent gets at most two resources.
		
		\begin{lemma}
			When $t=1$, for \efpashortE{} with any~$\Efficiency  \in \{\usw, \esw,  \allowbreak \size, \mcar\}$, there exists a desired envy-free allocation if and only if 
			\begin{enumerate}
				\item there exists a desired envy-free allocation where each agent gets exactly one resource, or
				\item there exists a desired envy-free allocation where some agents get one resource they like and the remaining agents get two resources.
			\end{enumerate} 
		\end{lemma}
		
		\begin{proof}
			According to~\Cref{lem:bivalues-equivalent}, it suffices to consider one efficiency measure, say $\esw$.
			The ``if'' direction is trivial and we show the ``only if'' direction.
			We define a new utility function $u'$ for each agent such that $u'_i(r_j)=u_i(r_j)-1$ for any resource $r_j \in \Resources$.
			Since the value of each resource is decreased by 1, it is easy to see that there exists an envy-free allocation where each agent gets exactly one resource under the original utility function $u$ if and only if there exists such an allocation under $u'$.
			Notice that $u'$ is a binary valuation, and according to \Cref{lem:NEB=1}, there exists such an allocation under $u'$ if and only if $|X| \leq |Y_L|$.
			Suppose $|X| > |Y_L|$, which means case (1) does not happen, we show that case (2) must happen.
			Since case (1) does not happen, for any envy-free allocation, at least one agent gets more than one resource.
			Moreover, since $|X_S|>|Y_S|$, we have $|Y| < 2|X|$.
			Since $u_i(r_j) \in \{1,2\}$, if one agent gets more than two resources, than all the other agents should get at least 2 resources to be envy-free, which overall needs more than $2|X| > |Y|=|R|$ resources and is impossible.
			Therefore, in any envy-free allocation, every agent gets either one resource or two resources.
			Moreover, every agent who gets one resource must get one resource they like, as otherwise they will envy agents who get two resources.
		\end{proof}
		
		Open Question: For any $\Efficiency  \in \{\usw, \esw, \size, \mcar\}$, is \efpashortE{} with bivalued valuations (1 and 2) and $t=1$ NP-hard?

	\end{document}